\documentclass{article}
\usepackage[utf8]{inputenc}
\usepackage[T1]{fontenc}
\usepackage{graphicx}
\usepackage{amsmath}
\usepackage{amssymb}
\usepackage{mathrsfs}
\usepackage{braket}
\usepackage{geometry}
\usepackage{here}
\usepackage{lmodern}
\usepackage{array}
\usepackage{multicol}
\title{Proposal for the formation of ultracold deeply-bound RbSr dipolar molecules by all-optical methods }
\author{A. Devolder$^1$ \and E. Luc-Koenig$^1$ \and O. Atabek$^2$ \and M. Desouter-Lecomte$^3$ \and O. Dulieu$^1$}
\date{$^1$ Laboratoire Aimé Cotton, CNRS, Université Paris-Sud, ENS Paris-Saclay, Université Paris-Saclay, B\^at. 505,91405 Orsay Cedex,France \\
	  $^2$ Institut des Sciences Moléculaires d'Orsay (ISMO, UMR8214) CNRS, Université Paris-Sud, Université Paris-Saclay, Bât. 520, 91405, Orsay, France \\
	  $^3$ Laboratoire de Chimie Physique (LCP,UMR 8000) CNRS, Université Paris-Sud, Université Paris-Saclay, Bât. 349,  91405, Orsay France}
		
\begin{document}
\maketitle
\begin{abstract}
Ultracold paramagnetic and polar diatomic molecules are among the promising systems for quantum simulation of lattice-spin models. Unfortunately, their experimental observation is still challenging. Based on our recent \textit{ab-initio} calculations, we analyze the feasibility of all-optical schemes for the formation of ultracold $^{87}$Rb$^{84}$Sr bosonic molecules. First, we have studied the formation by photoassociation followed by spontaneous emission. The photoassociation rates to levels belonging to electronic states converging to the $^{87}$Rb$(5s\,^2S)$+$^{84}$Sr($5s5p\,^3P_{0,1,2}$) asymptotes are particularly small close to the asymptote. The creation of molecules would be more interesting by using deeply levels that preferentially relaxes to the $v''=0$ level of the ground state. On the other hands, the photoassociation rates to levels belonging to electronic states converging to the Rb$(5p\,^2P_{1/2,3/2})$+Sr($5s^2\,^1S$) asymptotes have high value close to the asymptote. The relaxation from the levels close to the asymptotes creates weakly-bound molecules in mosty only one vibrational level. Second, stimulated Raman adiabatic passage (STIRAP) achieved in a tight optical trap efficiently creates weakly-bound ground-state molecules in a well-defined level, thus providing an alternative to magnetic Feshbach resonances to implement several schemes for an adiabatic population transfer toward the lowest ground-state level of RbSr. Finally, we have studied STIRAP process for transferring the weakly-bound molecules into the $v''=0$ level of the RbSr ground state. 
\end{abstract}

\section{Introduction}

Ultracold diatomic molecules, namely with translational motion cooled down to temperatures well below one millikelvin, and internal degrees of freedom reduced to a single quantum level \cite{bahns1996}, are nowadays well recognized as promising systems for quantum simulation, quantum computation, ultracold chemistry and precision measurements. This is particularly true for those species which possess additional internal properties like a permanent electric dipole moment (PEDM) in their own frame, and/or a magnetic dipole moment, as they can be manipulated by external electric and magnetic fields \cite{Review_cold_molecule_1,Review_cold_molecule_2,Quantum_comp_RbSr,polar_paramagnetic_molecules_field,Micheli_lattice_spin}.

The first translationally-ultracold molecular species ever produced were homonuclear, namely Cs$_2$ \cite{Cs2_Comparat}, and Rb$_2$ molecules \cite {gabbanini2000}. The formation process relied on photoassociation (PA) of ultracold atomic pairs followed by radiative emission (RE) down to the electronic ground state \cite{thorsheim1987}. Shortly after, several groups were able to create heteronuclear diatomic species with the same approach \cite{jones2006}. A first breakthrough came with the direct observation of ultracold molecules formation (UMF) in the lowest rovibrational level ($v=0, J=0$) of their ground state \cite{deiglmayr2008a}, with some ability of control of their internal state \cite{viteau2008,wakim2012,manai2012}. The fully-controlled creation of ultracold dipolar molecules was demonstrated at about the same time on the KRb polar species \cite{KRb,KRb_2,borsalino2014}, but using the alternative approach of magnetoassociation of an atom pair into a weakly-bound molecule, followed by a stimulated Raman adiabatic passage (STIRAP) to transfer population into the lowest energy level of the KRb electronic ground state. 

Ground-state species exhibiting an additional magnetic moment are for instance diatomic molecules composed of an alkali-metal atom and an alkaline-earth atom (or an Ytterbium atom), in which the magnetic dipole moment comes from the existence of an unpaired electron. Surprisingly, the spectroscopy of such diatomic molecules is still poorly known. The recent interest for such ultracold species triggered several investigations at relatively low resolution \cite{krois2013,krois2014,pototschnig2015,gerschmann2017,schwanke2017,ciamei2018}. However these species are still challenging to create in the ultracold domain. After an initial prediction \cite{zuchowski2010a}, magnetic Feshbach resonances have been observed for $^{87}$Rb$^{88}$Sr and $^{87}$Rb$^{87}$Sr molecules \cite{barbe2017}, but they are not yet used for the formation of weakly-bound ground-state $^{87}$Rb$^{88}$Sr .

All-optical methods are attractive to create ultracold molecules as they do not rely on peculiarities of the molecular structure like the presence of Feshbach resonances in the ground state at moderate magnetic fields. Here we model the PA+RE sequence mentioned above for the $^{87}$Rb$^{84}$Sr bosonic species. As expected we find that it is not selective enough to populate a single quantum level in the molecular ground state. Therefore we consider the motional levels of a tight trap \cite{jaksch2002} to implement a STIRAP transfer, as previously demonstrated with Sr$_2$ molecules \cite{STIRAP_production_Sr2_1,STIRAP_production_Sr2_2}. Our calculations are based on the RbSr electronic structure data previously obtained in our group \cite{RbSr_FCI_GS,Zuchowski_RbSr}, which are recalled in Section \ref{sec:structure}. We compute in Section \ref{sec:PA} PA and UMF rates, when the PA laser is tuned to the red of either the ($5\,^2S_{1/2} \rightarrow 5\,^2P_{1/2,3/2}$) resonant transitions in $^{87}$Rb, or the ($5\,^1S_{0} \rightarrow 5\,^3P_{0,1,2}$) intercombination transitions in $^{84}$Sr. The transition $^1$S$_0$ $\rightarrow$ $^3$P$_1$ is indeed employed for the cooling of Sr atoms in ongoing experiments for quantum degenerate mixtures of strontium and rubidium atoms \cite{Mixture_BEC_Sr_Rb}. Relying on the obtained knowledge about transition dipole moment, we propose promising candidate levels to implement STIRAP process in a tight optical trap to create weakly-bound ultracold $^{87}$Rb$^{84}$Sr ground state molecules (Section \ref{sec:STIRAPtrap}), and their transfer into the lowest vibrational level of their ground state using a second STIRAP sequence (Section \ref{sec:STIRAPgroundstate}). For convenience purpose in the calculations, atomic units of distance (1~a.u.$=a_0=0.052917721067$ $10^{-10}$ ~m), energy (1~a.u.=1 hartree=219474.6313702~cm$^{-1}$) and electric dipole moment (1~a.u.=2.54175~D) will be used throughout the paper, except otherwise stated.

\section{Electronic structure of the RbSr molecule}
\label{sec:structure}

\begin{figure}
	\centering
	\includegraphics[width=8cm]{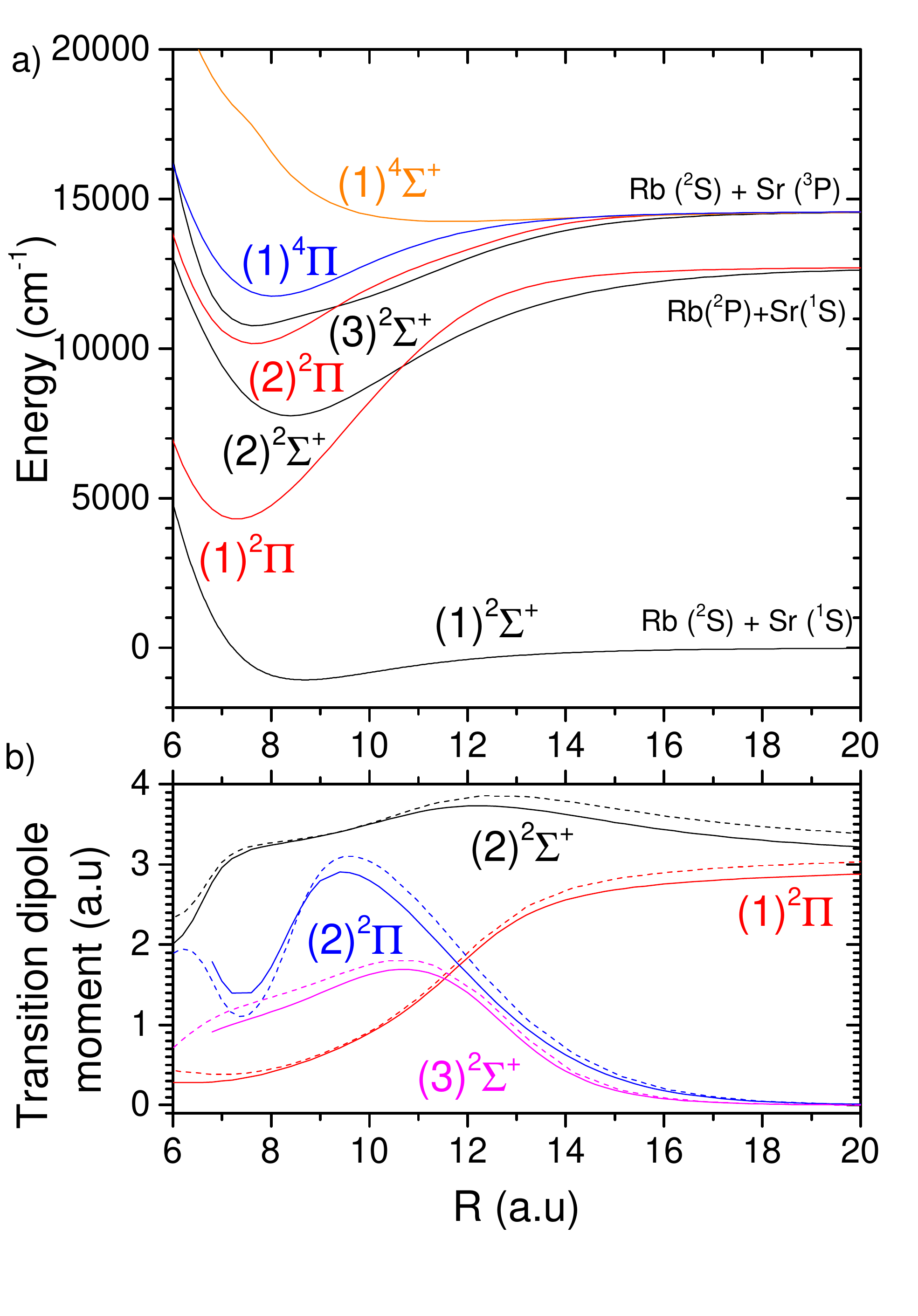}
	\caption{Upper panel: potential energy curves of the electronic states relevant for the present study, restricted to distances smaller than 20~a.u. for clarity. Lower panel: transition dipole moment between the $(1)^2\Sigma^+$ ground state and the $^2\Sigma^+$ and $^2\Pi$ states of the upper panel \cite{Zuchowski_RbSr} (solid curves), compared to those of Ref.\cite{pototschnig2014} (dashed lines).  We note that our computed transition dipole moments (TDMs) properly match the atomic TDMs at large distances, while those of Ref.\cite{pototschnig2014} are overestimated by about 10\% at this limit.}
	\label{fig:pot}
\end{figure}

\begin{figure}
	\centering
	\includegraphics[width=8cm]{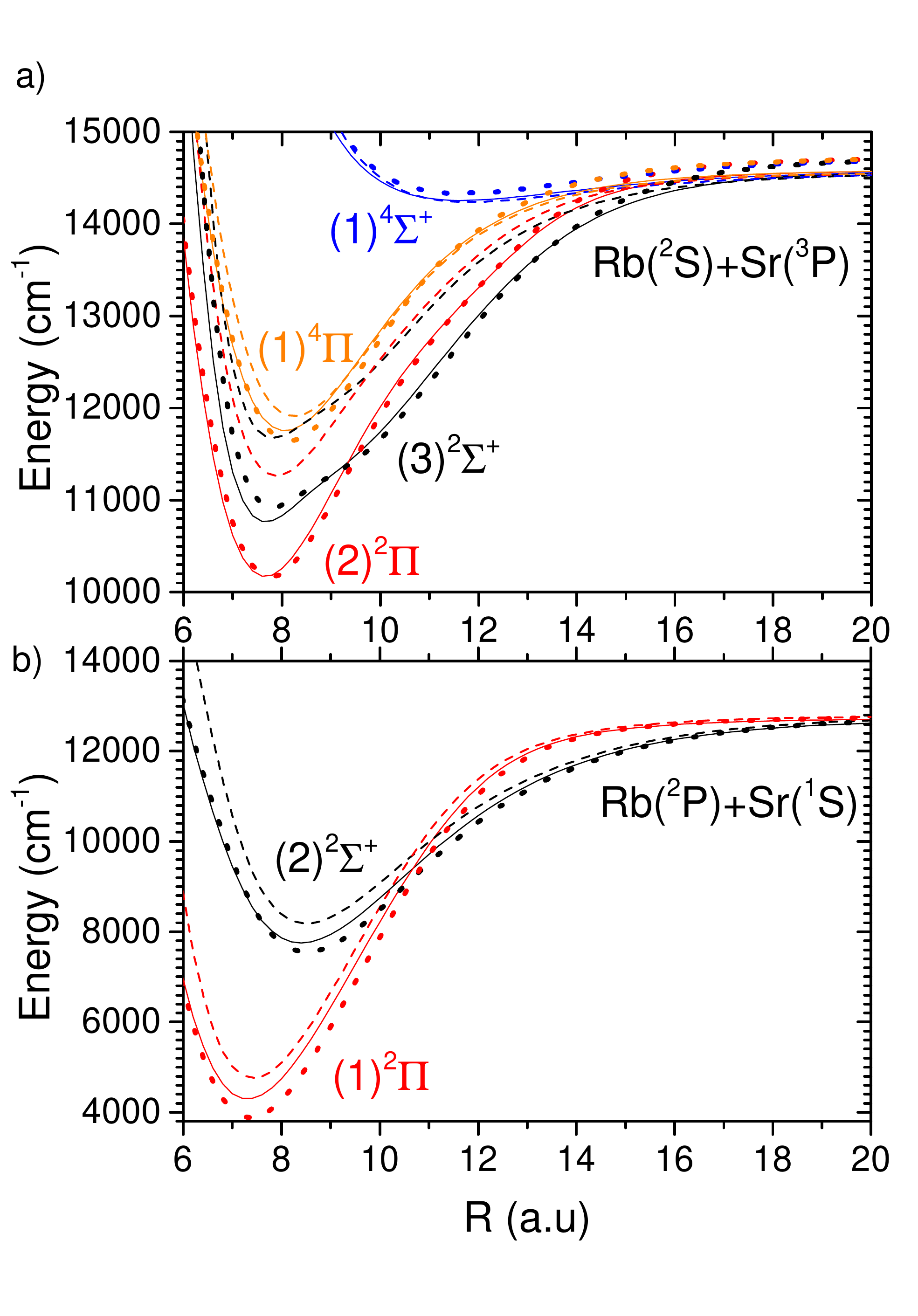}
	\caption{Comparison of potential energy curves of the lowest excited states of RbSr calculated with FCI ECP+CPP (solid lines), EOM-CC (dashed lines) \cite{Zuchowski_RbSr}, and MCSCF-MRCI (dotted lines) methods \cite{RbSr_Ernst}.} 
	\label{fig:PEC-comparaison}
\end{figure}

In this work, we are interested in the states correlated to the three lowest dissociation limits of RbSr (turning into six limits when spin-orbit interaction is included), listed with increasing energy: Rb $(5s\,^2S)$ + Sr ($5s^2\,^1S$), Rb $(5p\,^2P)$ + Sr ($5s^2\,^1S$) and Rb $(5s\,^2S)$ + Sr ($5s5p\,^3P$). They give rise to three sets of electronic states (labeled in Hund's case a notation $(N)^{2S+1}\Lambda$), namely ($(1)^2\Sigma^+$, or more commonly $X^2\Sigma^+$), ($(2)^2\Sigma^+, (1)^2\Pi$), and ($(3)^2\Sigma^+, (1)^2\Pi, (1)^4\Sigma^+, (1)^4\Pi$), respectively. The corresponding potential energy curves (PECs), and transition dipole moments (TDMs) between the $X^2\Sigma^+$ ground state and several excited electronic states are displayed in Fig. \ref{fig:pot}. For the calculations in the next sections, we have selected full configuration-interaction (FCI) calculation performed on the three valence electrons moving in the field of relativistic large effective core potentials (ECPs), including core-polarization potentials (CPP), and extrapolated to large distances, which were reported in our previous work \cite{Zuchowski_RbSr}. 
Several features of the PECs are important to notice for the following. First, except for the (1)$^4\Sigma^+$ state, the equilibrium distances of excited-state PECs are significantly smaller than the one of the ground-state PEC. Second, two  curve crossings are visible, between the (2)$^2\Sigma^+$ and (1)$^2\Pi$ PECs, and between the (3)$^2\Sigma^+$ and (2)$^2\Pi$ PECs. These features are also present in two other available calculations displayed in Fig. \ref{fig:PEC-comparaison} from very different methods, namely the EOM-CCSD (equation-of-motion coupled-cluster method limited to singly and doubly excited configurations) method employed in Ref.\cite{Zuchowski_RbSr,RbSr_Ernst}, and the MCSCF-MRCI (MultiConfigurational Self-Consistent Field-Multi-Reference Configuration Interaction) method of Ref.\cite{RbSr_Ernst}. Figure \ref{fig:PEC-comparaison} reveals a good overall agreement among all the results, recalling however that the EOM-CCSD results for the excited electronic states are probably less accurate than the other results, as already discussed in Ref.\cite{Zuchowski_RbSr}. We also note in the upper panel, that the PECs from Ref.\cite{RbSr_Ernst} converge to a dissociation energy larger by 107~cm$^{-1}$  than ours. This is related to the excitation energy of the Sr($5s5p\,^3P$) level, found 20~cm$^{-1}$ above (resp. 87~cm$^{-1}$ below) the experimental one, in the  MCSCF-MRCI calculations (resp. FCI and EOM-CCSD calculations \cite{Zuchowski_RbSr,guerout2010}). In order to illustrate in a complementary way the above results, we display in Table \ref{tab:Spectro_param} the spectroscopic constants of these PECs, as well as the coefficient $C_6$ of the leading-order term of the long-range van der Waals interaction between Rb and Sr \cite{jiang2013}. For the ground state, the calculations of Chen \textit{et al.}  \cite{Chen_RbSr} is also included. Deeply-bound spectra with thermoluminescence and Laser induced fluorescence between (X)$^2\Sigma^+$ and (2)$^2\Sigma^+$ states have been made by A. Ciamei et al. \cite{ciamei2018}. They simulated fluorescence spectra using PECs from FCI ECP+CPP, EOM-CCSD and MSCF-MRC calculations and compared with the experimental one. For all three calculations, only few experimental band heads can be identified unambiguously. The predicted wavenumber for the 0-0 band head is very close in the case of the MSCF-MRCI calculation. The difference with the FCI ECP+CPP and EOM-CCSD calculation are respectively 25 cm$^{-1}$ and 400 cm$^{-1}$. On the other hands, the FCI-ECP and EOM CCSD calculations can predict more bands than the MSCF-MRCI calculation and the shape of the simulated spectrum is closer to the experimental one. 

\begin{table}
	\centering
	\caption{Main spectroscopic constants of the $^{87}$Rb$^{84}$Sr electronic states correlated to the three lowest dissociation limits. The labels ''a'' and  ''b'' for Ref.\cite{Zuchowski_RbSr} refer to the calculations with the FCI-ECP method, and the EOM-CCSD method, respectively (see text).}
	\begin{tabular}{ccccccc}
 \hline \hline
 State           & $R_e$ & $D_e$     &$\omega_e$ &$C_6$ \cite{jiang2013}&Limit& \\
                 &($a_0$)&(cm$^{-1}$)&(cm$^{-1}$)&(a.u.)&&\\  \hline
 (X)$^2\Sigma^+$ &8.69   &1073.3     &38.98      &3699 &Rb$(5s\,^2S)$+Sr($5s^2\,^1S$)&\cite{Zuchowski_RbSr}a\\
                 &8.82   &1040.5     &38.09      &     &                             &\cite{Zuchowski_RbSr}b\\
								 &8.65   &1283.5     &42.1       &     &                             &\cite{RbSr_Ernst}\\
								 &8.827  &1017.58    &36.017     &     &                             &\cite{Chen_RbSr}\\
 (2)$^2\Sigma^+$ &8.40   &4982.9     &58.37      &23324&Rb$(5p\,^2P)$+Sr($5s^2\,^1S$)&\cite{Zuchowski_RbSr}a\\
                 &8.51   &4609.6     &60.20      &     &                             &\cite{Zuchowski_RbSr}b\\
                 &8.54   &5144.3     &58.9       &     &                             &\cite{RbSr_Ernst}\\
 (1)$^2\Pi$      &7.31   &8439.8     &79.50      &8436 &                             &\cite{Zuchowski_RbSr}a\\
                 &7.42   &8038.6     &83.19      &     &                             &\cite{Zuchowski_RbSr}b\\
                 &7.39   &8770.2     &79.5       &     &                             &\cite{RbSr_Ernst}\\
 (3)$^2\Sigma^+$ &7.67   &3828.0     &65.26      &8929 &Rb$(5s\,^2S)$+Sr($5s5p\,^3P$)&\cite{Zuchowski_RbSr}a\\
                 &7.81   &2892.4     &62.48      &     &                             &\cite{Zuchowski_RbSr}b\\
                 &7.84   &3677.8     &57.4       &     &                             &\cite{RbSr_Ernst}\\
 (2)$^2\Pi$      &7.65   &4421.2     &67.60      &5716 &                             &\cite{Zuchowski_RbSr}a\\
                 &7.88   &3303.5     &63.37      &     &                             &\cite{Zuchowski_RbSr}b\\      
                 &7.80   &4450.3     &65.8       &     &                             &\cite{RbSr_Ernst}\\   
 (1)$^4\Sigma^+$ &11.63  &336.3      &15.42      &8929 &                             &\cite{Zuchowski_RbSr}a\\
                 &11.81  &329.2      &15.03      &     &                             &\cite{Zuchowski_RbSr}b\\
                 &11.64  &396.7      &16.        &     &                             &\cite{RbSr_Ernst}\\
 (1)$^4\Pi$      &8.06   &2838.1     &56.98      &5716 &                             &\cite{Zuchowski_RbSr}a\\ 
                 &8.24   &2655.7     &54.95      &     &                             &\cite{Zuchowski_RbSr}b\\ 
                 &8.16   &3053.9     &57.6       &     &                             &\cite{RbSr_Ernst}\\ \hline \hline
	\end{tabular}
	\label{tab:Spectro_param}
\end{table}

It is well known that a model for a PA spectrum depends on two crucial inputs: the long-range behavior of the PECs of the relevant states, and the scattering length of the ground state. Therefore the (X)$^2\Sigma^+$ ground state PEC has been smoothly matched at 15$a_0$ to an asymptotic expansion expressed as $-C_6/R^6-C_8/R^8-C_{10}/R^{10}$, with $C_8=4.609\times 10^{5}$~a.u. and $C_{10}=5.833\times 10^7$~a.u. \cite{jiang2013}.  In contrast, the long-range expansion of the excited-state PECs has been restricted to the $-C_6/R^{-6}$ term (Table \ref{tab:Spectro_param}). 

In the absence of a global PEC determined spectroscopically for the ground state, one cannot rely on the scattering length provided by the computed PEC. However the binding energies of the two uppermost vibrational levels of the $^{87}$Rb$^{84}$Sr molecule relative to the Rb $(5s\,^2S, F=1)$ + Sr ($5s^2\,^1S$) limit (where $F$ denotes the total angular momentum of the Rb atom accounting for the nuclear spin) have been recently measured by A. Ciamei and coworkers \cite{ciamei2018} in a two-photon photoassociation experiment leading to $9.67\times10^{-4}$~cm$^{-1}$ and $2.49$~cm$^{-1}$. Therefore we have slightly modified the position of the repulsive wall of the PEC around the dissociation limit  in order to match these experimental energies. The calculations of ground state eigenenergies were performed with the Mapped Fourier Grid Hamiltonian method (MFGH) \cite{MFGH_1,MFGH_2,MFGH_3,MFGH_4}. The adjustment was constrained to the condition that the spectroscopic data of Table~\ref{tab:Spectro_param} remain unchanged (i.e. the bottom of the PEC is unchanged). The best agreement was found when moving the inner turning point by 0.042~$a_0$ toward smaller distances. After this adjustment, the scattering length has a value of 89.3 a.u. which is close to the experimental one (92.7 a.u.) \cite{ciamei2018} .

Important features of TDMs can be pointed out, and may have a strong influence on the optical response of RbSr molecules. First, the spin selection rule forbids transitions between doublet and quartet states (the latter are not reported in Fig.\ref{fig:pot}). Second, the atomic transition Rb $(5s\,^2S)$ $\rightarrow$ Rb $(5p\,^2P)$ is allowed while the atomic transition Sr ($5s^2\,^1S$) $\rightarrow$ Sr ($5s^2\,^3P$) is spin-forbidden. This atomic selection rule is visible in the long-range part of the TDMs on Fig.\ref{fig:pot} b). Finally, at short-range, the TDM with the (2) $^2\Sigma^+$ state is  large, while the TDM with the (1) $^2\Pi$ state is close to zero at short range. 

Spin-orbit (SO) splitting being large for the lowest excited states of both atoms ($\Delta_{\textrm{fs}}^{Rb}=237.1$~cm$^{-1}$ and $\Delta_{\textrm{fs}}^{Sr}=581.1$~cm$^{-1}$), it must be taken into account to model the PA spectrum. We used the same approach as in Ref.\cite{Zuchowski_RbSr}: first the two sets of PECs correlated to the Rb $(5p\,^2P)$ + Sr ($5s^2\,^1S$) and Rb $(5s\,^2S)$ + Sr ($5s5p\,^3P$) dissociation limits are considered independently, and the atomic SO operators $\hat{W}_{\textrm{so}}^{Rb}=A^{Rb}\vec{\ell}^{Rb} . \vec{s}^{Rb}$ and $\hat{W}_{\textrm{so}}^{Sr}=A^{Sr}(\vec{\ell}_1^{Sr} . \vec{s}_1^{Sr}+\vec{\ell}_2^{Sr} . \vec{s}_2^{Sr}$), respectively, are used as perturbations to the Hamiltonian containing the kinetic operator and the electrostatic interactions. The states including SO are labeled according to the projection $|\Omega|$ of the total electronic angular momentum on the molecular axis (Hund case (c)). For the former asymptote, the $|\Omega|=3/2$ Hamiltonian matrix (including electrostatic interaction and SO) reduces to a single element $W_{\textrm{so}}^{Rb}(|\Omega|=3/2)= V((1)^2\Pi)+2 A^{Rb}$, where $A^{Rb}=\Delta_{\textrm{fs}}^{Rb}/3$, while the matrix for $|\Omega|=1/2$ reads
\begin{equation}
W_{\textrm{so}}^{Rb}\left(|\Omega|=\frac{1}{2}\right)=\begin{pmatrix}
V((2)^2\Sigma^+) & \sqrt{2} A^{Rb} \\
\sqrt{2} A^{Rb} & V((1)^2\Pi)+A^{Rb}
\end{pmatrix}.
\label{eq:Rb-omega12}
\end{equation}
For the latter dissociation limit, defining $A^{Sr}=\Delta_{\textrm{fs}}^{Sr}/3$, the maximal value of $|\Omega|$ is 5/2, with a single matrix element  $W_{\textrm{so}}^{Sr}(|\Omega|=5/2)= V((1)^4\Pi)+A^{Sr}$, while the matrices are, for $|\Omega|=3/2$ and $|\Omega|=1/2$ respectively,
\begin{equation}
W_{\textrm{so}}^{Sr}\left(|\Omega|=\frac{3}{2}\right)=\begin{pmatrix}
V((2)^2\Pi)+\frac{2}{3} A^{Sr} & \sqrt{\frac{1}{3}}A^{Sr} & -\frac{\sqrt{2}}{3} A^{Sr} \\
\sqrt{\frac{1}{3}} A^{Sr} & V((1)^4\Sigma^+) & \sqrt{\frac{2}{3}} A^{Sr} \\
-\frac{\sqrt{2}}{3} A^{Sr} & \sqrt{\frac{2}{3}} A^{Sr} & V((1)^4\Pi)+\frac{1}{3} A^{Sr}
\end{pmatrix}
\label{eq:Sr-omega12}
\end{equation}
and 
\begin{equation}
W_{\textrm{so}}^{Sr}\left(|\Omega|=\frac{1}{2}\right)=\begin{pmatrix}
V((3)^2\Sigma^+) & \sqrt{\frac{8}{9}}A^{Sr} & 0 & -\frac{1}{3} A^{Sr} & \sqrt{\frac{1}{3}}A^{Sr} \\
\sqrt{\frac{8}{9}}A^{Sr} & V((2)^2\Pi)-\frac{2}{3}A^{Sr} & \frac{1}{3} A^{Sr} & -\frac{\sqrt{2}}{3} A^{Sr} & 0 \\
0 & \frac{1}{3} A^{Sr} & V(^4\Sigma^+) & \sqrt{\frac{8}{9}}A^{Sr} & \sqrt{\frac{2}{3}}A^{Sr} \\
-\frac{1}{3} A^{Sr} & -\frac{\sqrt{2}}{3} A^{Sr} & \sqrt{\frac{8}{9}}A^{Sr} & V((1)^4\Pi)-\frac{1}{3} A^{Sr} & 0 \\
\sqrt{\frac{1}{3}} & 0 & \sqrt{\frac{2}{3}}A^{Sr} & 0 & V((1)^4\Pi)-A^{Sr}
\end{pmatrix}
\label{eq:Sr-omega32}
\end{equation} 

\begin{figure}
	\centering
	\includegraphics[width=16cm]{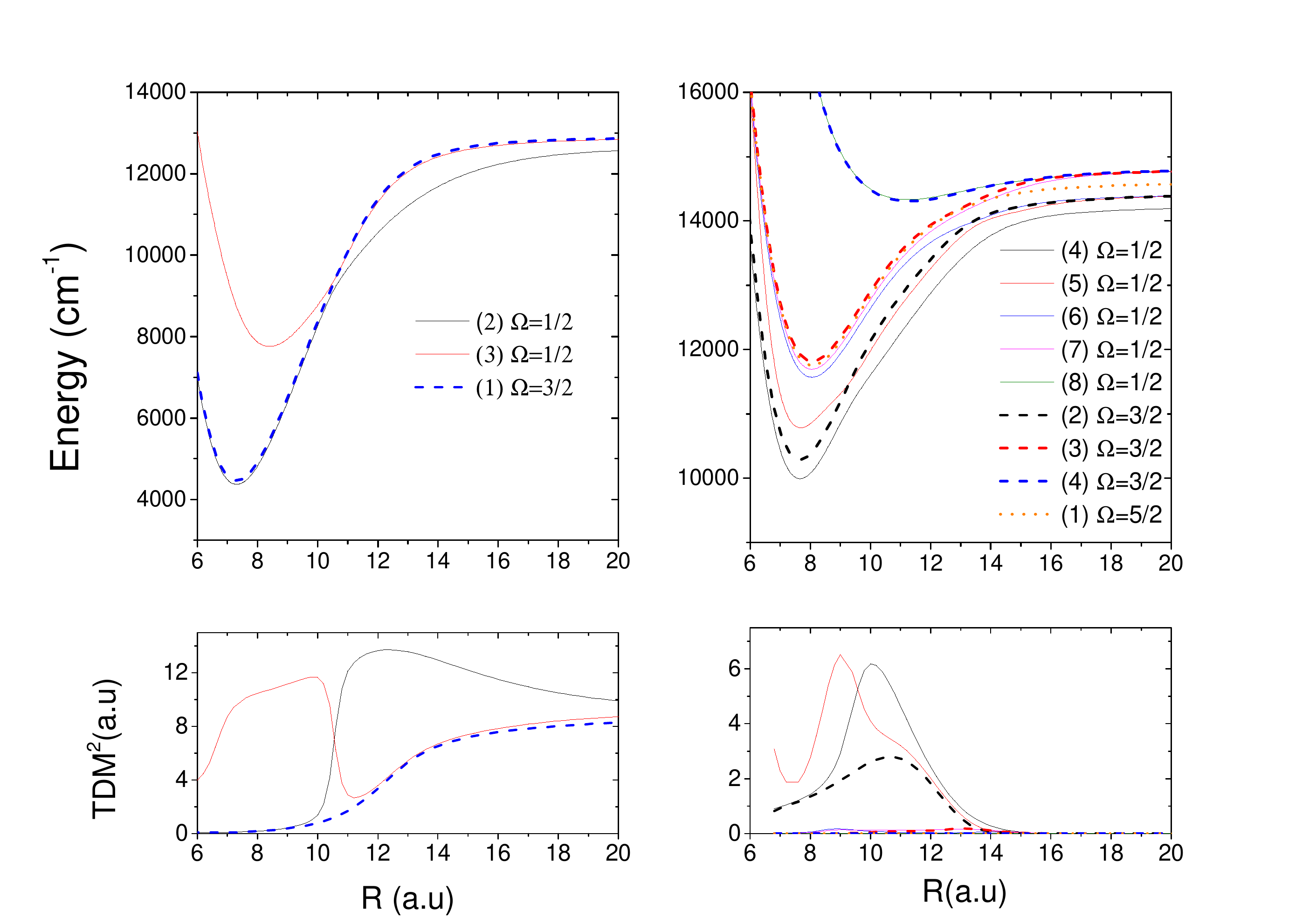}
	\caption{RbSr potential energy curves of excited states including spin-orbit interaction as described in the text, and corresponding effective squared transition dipole moments from the $X^2\Sigma^+$ ground state towards the states correlated to Rb$(5p\,^2P_{1/2,3/2})$+Sr($5s^2\,^1S$) (left column), and to Rb$(5s\,^2S)$+Sr($5s5p\,^3P_{0,1,2}$) (right column).}
	\label{fig:PEC_SO}
\end{figure}

The Hund's case c PECs (N)$\Omega$ including SO interaction are straightforwardly obtained by diagonalization of the full hamiltonian involving these matrices at each fixed R value (Fig.\ref{fig:PEC_SO}). The asymptote Rb $(5p\,^2P)$ + Sr ($5s^2\,^1S$) is split in Rb $(5p\,^2P_{1/2})$ + Sr ($5s^2\,^1S$) and Rb $(5p\,^2P_{3/2})$ + Sr ($5s^2\,^1S$) while the asymptote Rb $(5s\,^2S)$ + Sr ($5s5p\,^3P$) is split in Rb $(5s\,^2S)$ + Sr ($5s5p\,^3P_0$), Rb $(5s\,^2S)$ + Sr ($5s5p\,^3P_1$) and Rb $(5s\,^2S)$ + Sr ($5s5p\,^3P_2$) (hereafter referred to as the  $^3P_0$,  $^3P_1$ and $^3P_2$ asymptotes, in short). A single $\Omega=1/2$ PEC is correlated to the $^3P_0$ asymptote, while two such PECs match  the $^3P_1$ and $^3P_2$ asymptotes. Similarly, a single $\Omega=3/2$ PEC is correlated to the $^3P_1$ asymptote and two PECs to the  $^3P_2$ asymptote. The equilibrium distances for excited states PECs are still smaller than for the ground state, except for the $(8) 1/2$ and $(3) 3/2$ states composed mainly of (1)$^4\Sigma^+$ state. The crossings in the Hund case (a) PECs become avoide crossings in the Hund case (c).  

We display in Table \ref{Spectro_param_comparaison_SO} the corresponding fundamental spectroscopic constants, as they are provided in several other publications \cite{RbSr_Ernst,Chen_RbSr}. As expected from Fig.\ref{fig:PEC-comparaison}, the equilibrium distances $R_e$ and the harmonic constants $\omega_e$ are those of the states without SO, as the avoided crossings occur far from $R_e$. The dissociation energies are significantly changed, reflecting the magnitude of the atomic SO splittings. As already noted, our results are in good agreement with those of Ref. \cite{RbSr_Ernst}. In contrast, significant differences are found with the work of Ref. \cite{Chen_RbSr}, in particular for the well depth and for the $\omega_e$ constant. In the latter work, the authors used the relativistic Dirac-Coulomb Hamiltonian where the electronic spin and consequently the related $R$-dependent relativistic interactions are explicitly accounted for in the Hamiltonian under a four-component framework, as initially developed in the approach of Ref.\cite{sorensen2009}, so that the spin-orbit interaction is included in a non-perturbative way. But they used a basis set which is significantly smaller than the one of Ref.\cite{RbSr_Ernst}, which thus may not be fully appropriate for excited states. Such differences in the PECs may also indicate a noticeable variation of the molecular SO coupling with the internuclear distance. Note that we have proved for the heaviest alkali atom Fr \cite{aymar2006} that an electronic structure calculation including only the scalar relativistic term (as performed here) yields satisfactory electronic atomic orbitals even for such a heavy species.  

In Fig.\ref{fig:PEC_SO} we also displayed the $R$-dependent quantity labeled with effective $\mathrm{TDM}^2$, representing, for each molecular state $(N)\Omega$ including SO, the linear combination of the squared $R$-dependent TDMs of Fig.\ref{fig:pot} associated to the transitions from the ground state toward the states involved in $(N)\Omega$. The weights of this combination are the squared components of the eigenvector associated to $(N)\Omega$. It should be noted that this quantity is not the one to consider for actually computing transition probabilities, \textit{i.e.} it should not be used in an integration over $R$ weighted by a pair of radial vibrational wave functions. Therefore the $\mathrm{TDM}^2$ quantities, referred to as ''effective squared TDM'' for convenience, only provide a qualitative description of the actual mixture of states without SO composing $(N)\Omega$, by comparison between Fig.\ref{fig:PEC_SO} and Fig.\ref{fig:pot}. As expected, our results for the $\mathrm{TDM}^2$ quantities are very similar to those of Ref.\cite{RbSr_Ernst}. This representation of the TDMs allows us to point some features that have an impact on PA. First, the avoided crossings between PECs induce crossings between TDM curves due to the change of the nature of the electronic states. At short range, the (2) $1/2$ state (resp. the (3) $1/2$ state) is mainly composed of (2) $^2\Pi$ state (resp. (1) $^2\Sigma^+$ state). This results in a high value for the former, and a low value for the latter. The $^2\Pi$ state composition of the (1) $\Omega=3/2$ state is also clear. For the TDMs for the states correlated to Rb$(5s\,^2S)$+Sr($5s5p\,^3P_{0,1,2}$), the main characteristic is the non-zero value for (6),(7) and (8) $\Omega=1/2$ states. Even if they are constituted of quartet state at short-range, the mixture with doublet states due to the spin-orbit interaction implies non-zero values at larger distance. At short-range, we can also notice that the (4) $\Omega=1/2$ and (2) $\Omega=3/2$ states are mainly composed by (2) $^2\Pi$ state while the (5) $\Omega=1/2$ state is mainly constituted by (3) $^2\Sigma^+$ state.  

\begin{table}
	\caption{Main spectroscopic constants of the $^{87}$Rb$^{84}$Sr electronic states including spin-orbit interaction, as compared to other published data. The corresponding dissociation limits are also indicated for clarity sake.}
	\resizebox{17cm}{3.3cm}{
		\centering
		\begin{tabular}{c|ccc|ccc|ccc|c}
		\hline
		\hline
		& \multicolumn{3}{c}{FCI ECP+CPP \cite{Zuchowski_RbSr} } & \multicolumn{3}{c}{KR-MRCI \cite{Chen_RbSr}} & \multicolumn{3}{c}{MCSCF-MRCI \cite{RbSr_Ernst}}& Asymptotes  \\
		State & $R_e$ ($a_0$) & $D_e$ ($cm^{-1}$) & $\omega_e$ ($cm^{-1}$) & $R_e$ ($a_0$) & $D_e$ ($cm^{-1}$) & $\omega_e$ ($cm^{-1}$) & $R_e$ ($a_0$) & $D_e$ ($cm^{-1}$) & $\omega_e$ ($cm^{-1}$) \\
		\hline
		(2) $\Omega=\frac{1}{2}$ &7.30 &8283.9 &80.12&7.27&7883.09&85.73&7.43&8569.0&79.6&Rb$(5p\,^2P_{1/2})$+Sr($5s^2\,^1S$) \\
		(3) $\Omega=\frac{1}{2}$ &8.39 &5136.41&59.04&8.39&4683.56&58.43&8.56&5252.3&58.7&Rb$(5p\,^2P_{3/2})$+Sr($5s^2\,^1S$) \\
		(1) $\Omega=\frac{3}{2}$ &7.31 &8439.8 &79.50&7.29&7957.31&87.18&7.41&8727.6&80.4&\\
		(4) $\Omega=\frac{1}{2}$ &7.66 &4234.40&68.85&    &       &     &7.82&4202.3&64.1&Rb$(5s\,^2S)$+Sr($5s5p\,^3P_0$)\\
		(5) $\Omega=\frac{1}{2}$ &7.69 &3635.19&65.77&    &       &     &7.94&3436.2&52.4&Rb$(5s\,^2S)$+Sr($5s5p\,^3P_1$)\\
		(6) $\Omega=\frac{1}{2}$ &8.06 &2851.68&57.41&    &       &     &    &      &    &\\
		(7) $\Omega=\frac{1}{2}$ &8.03 &3112.45&57.63&    &       &     &    &      &    &Rb$(5s\,^2S)$+Sr($5s5p\,^3P_2$)\\
		(8) $\Omega=\frac{1}{2}$ &11.27&476.41 &20.57&    &       &     &    &      &    &\\
		(2) $\Omega=\frac{3}{2}$ &7.66 &4129.30&68.34&    &       &     &7.82&4107.4&64.5&Rb$(5s\,^2S)$+Sr($5s5p\,^3P_1$)\\
		(3) $\Omega=\frac{3}{2}$ &8.06 &2990.04&57.64&    &       &     &    &      &    &Rb$(5s\,^2S)$+Sr($5s5p\,^3P_2$)\\
		(4) $\Omega=\frac{3}{2}$ &11.30&501.73 &21.05&    &       &     &    &      &    &\\
		(1) $\Omega=\frac{5}{2}$ &8.06 &2990.04&56.98&    &       &     &    &      &    &\\
		\hline
		\hline
	\end{tabular}
	}
	\label{Spectro_param_comparaison_SO}
\end{table}

\section{Photoassociation of $^{87}$Rb$^{84}$Sr molecules}
\label{sec:PA}

\subsection{Methodology}

The photoassociation rate toward a vibrational level $v'$ of an excited electronic state $e$, at low laser intensity $I$, of a $^{87}$Rb$^{84}$Sr pair colliding in the the $X$ ground state with relative energy $E=k_BT$, is computed according to the perturbative approach reported in Ref. \cite{Pillet_Photoassociation}: 
\begin{equation}
R_{PA}(X \rightarrow e, v';T,I)=\frac{2\pi^{1/2}h}{c} \left(\frac{3}{2}\right)^{3/2} \lambda_{th} I A|\braket{\phi_{v'}^e|\mu_{eX}^{el}|u_{\ell}^X(E)}|^2 ,
\label{eq:pa-rate}
\end{equation}
where $\lambda_{th}=\sqrt{\frac{h^2}{3\mu k_BT}}$ is the de Broglie thermal wavelength of the atom pair with reduced mass $\mu$, $I$ is the intensity of photoassociation laser, $A$ is an angular factor, $\phi_{v'}^e(R)$ is the vibrational wave function of the photoassociated level $v'$, $u_{\ell}^X(k_BT,R)$ is the continuum wave function of the colliding pair assuming a rotational quantum number (partial wave) $\ell$, and $\mu_{eg}^{el}(R)$ is the $R$-dependent transition dipole moment between the X and $e$ electronic states. For simplicity, we have considered an $s$-wave collisional regime ($\ell=0$), and we have taken $A=1$, thus ignoring the dependence on the light polarization and on the internal states of the colliding atoms. The values of $I=10$~W/cm$^2$ and $T=5.5$~$\mu$K are typical of the ongoing experiment \cite{ciamei2018}, and are within the limits of validity of Eq. (\ref{eq:pa-rate}), namely in the linear regime for $I$, and  a non-degenerate quantum gas.

The radial wave functions above are calculated with the Mapped Fourier Grid Hamiltonian (MFGH) method \cite{MFGH_1}\cite{MFGH_2}\cite{MFGH_3}\cite{MFGH_4}, using a grid extending from $R_{min}=5 a_0$ to $R_{max}=2000 a_0$, containing up to 1551 points. We have checked the convergence of the calculations with the size of the grid. The continuum and vibrational levels of the $X^2\Sigma^+$ ground state are described in a single-channel representation, the wave functions (normalized to unity) discretized levels of the continuum being renormalized in energy at the end of the calculation by dividing the wave function by the square root of the level density \cite{luc-koenig2004}. The vibrational wave functions of the $^{87}$Rb$^{84}$Sr excited electronic states coupled by SO interaction are obtained from a multichannel representation according to Equations (\ref{eq:Rb-omega12}-\ref{eq:Sr-omega32}), and are therefore linear combinations of the related Hund's case a electronic states weighted by the radial wave functions $\phi_{v'^2\Sigma}^e$ and $\phi_{v'^2\Pi}^e$. The squared matrix elements of the transition dipole moment $\mu_{eg}^{el}(R)$ involve the contributions of the $^2\Sigma^+$ and $^2\Pi$ components of the coupled excited electronic states as
\begin{equation}
 |\braket{\phi_{v'}^e|\mu_{eg}|u_{0}^X(E)}|^2=|\braket{\phi_{v'^2\Sigma}^e|\mu_{^2\Sigma X}|u_{0}^X(E)}|^2+|\braket{\phi_{v'^2\Pi}^e|\mu_{^2\Pi X}|u_{0}^X(E)}|^2,
\label{eq:squared-tdm}
 \end{equation}
where the TDM functions $\mu_{^2\Sigma X}(R)$ and $\mu_{^2\Pi X}(R)$ are those displayed in Fig.\ref{fig:pot}. This equation is valid in the case of unpolarized light in the laboratory frame.
%
%

Two cases are of relevance for our study. First, the Sr intercombination (spin-forbidden) transition is used in ongoing experiments devoted to the formation of the quantum degenerate mixture of strontium and rubidium atoms \cite{Mixture_BEC_Sr_Rb}. Therefore we have studied the photoassociation close to the $^{87}$Rb$(5s\,^2S)$+$^{84}$Sr($5s5p\,^3P_{0,1,2}$) asymptotes. However PA will proceed in a very different way if it is implemented for laser frequencies close to the Rb dipole-allowed transition, namely, exploring bound levels close to the $^{87}$Rb$(5p\,^2P_{1/2,3/2})$+$^{84}$Sr($5s^2\,^1S$) asymptotes, as investigated theoretically in Ref. \cite{Chen_RbSr}. 

\begin{figure}
		\centering
		\includegraphics[width=12cm]{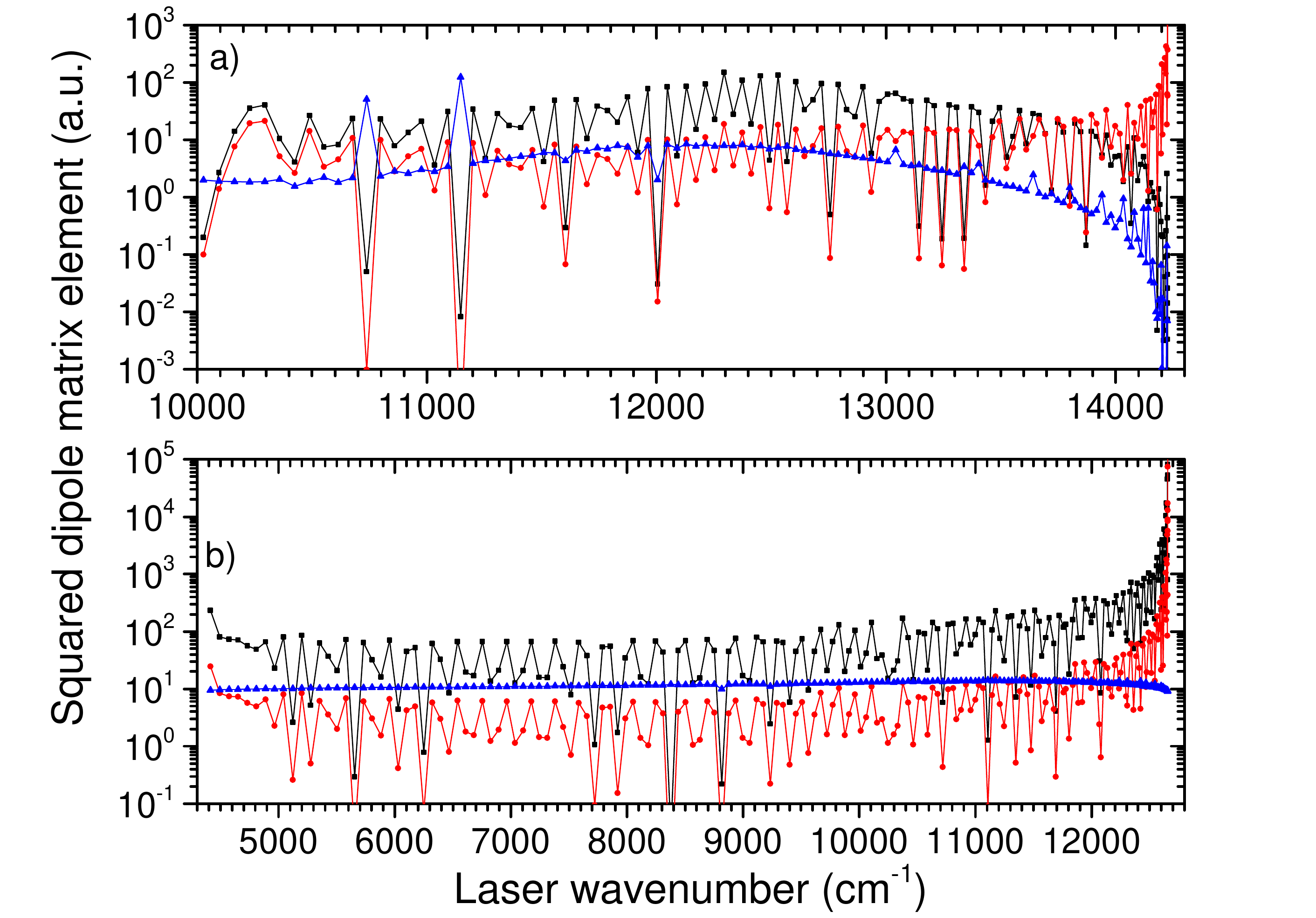}
		\caption{Squared dipole matrix element for (a) the (4) $\frac{1}{2}$ levels correlated to $^{87}$Rb$(5s\,^2S)$+$^{84}$Sr($5s5p\,^3P_0$), and (b) the (2) $\frac{1}{2}$ levels correlated to $^{87}$Rb$(5p\,^2P_{1/2})$+$^{84}$Sr($5s^2\,^1S$), as a function of the PA laser wavenumber. Black lines: squared TDM from Eq.(\ref{eq:squared-tdm}). Red lines: squared overlap (FCF) between the radial wave functions involved in the PA rate. Blue lines: value of squared TDM at the outer turning point $R_C$ of the radial wave function of the photoassociated level.}
		\label{fig:squaredTDME}
\end{figure}

Before the presentation of our systematic results for these situations, it is worthwhile to emphasize on the importance of accounting for the $R$ dependence of the TDM functions in the PA rates, which is not considered in Ref. \cite{Chen_RbSr}. For this purpose it is convenient to decompose the squared TDMs of Eq. (\ref{eq:squared-tdm}) as the product of the squared overlap between radial wave functions (or Franck-Condon factors, FCF), and the values of electronic transition dipole moment at the outer turning point $R_C$ of $\phi_{v'}^e$ for the corresponding excited electronic state :
\begin{equation}
|\braket{\phi_{v'}^e|\mu_{eg}|u_{0}^X(E)}|^2=\mu_{eg}^2(R_C)|\braket{\phi_{v'}^e|u_{0}^X(E)}|^2.
\label{eq:FC}
\end{equation}
The general trend of the squared TDMs is illustrated in Fig.\ref{fig:squaredTDME} with the (4) $\frac{1}{2}$ and the (2) $\frac{1}{2}$ states, respectively correlated to $^{87}$Rb$(5s\,^2S)$+$^{84}$Sr($5s5p\,^3P_{0}$ and $^{87}$Rb$(5p\,^2P_{1/2})$+$^{84}$Sr($5s^2\,^1S$). The squared TDMs have similar magnitude over most of the energy range of the potential wells, like the corresponding transition dipole moments in the molecular range ($R<12$~a.u. typically, see Fig. \ref{fig:pot}). Their behaviors are very similar to the FCF's ones. Close to the $^{87}$Rb$(5s\,^2S)$+$^{84}$Sr($5s5p\,^3P_{0}$) asymptote, the squared TDM vanishes as $\mu_{eg}^2(R_C)$ while the FCFs increase. Therefore the FCFs are not anymore representative of the behavior of the squared TDMs. This is in striking contrast with PA close to the $^{87}$Rb$(5p\,^2P_{1/2})$+$^{84}$Sr($5s^2\,^1S$) limit, for which the squared TDMs and the squared FCFs have similar behaviors. This decomposition of squared TDMs will also be useful for analyzing the PA spectra in the next section. 

In this Section we consider the formation of ultracold molecules by RE of the photoassociated molecules. For simplicity we assume, like in Ref.\cite{Pillet_Photoassociation}, that the RE probability is given by $\sum_{v''}|\braket{\phi_{v'}^e(R)|\phi_{v''}^g(R)}|^2$, such that the resulting ultracold molecule formation (UMF) rate per atom is written as
\begin{equation}
R_{mol}^{v'}=R_{PA}^{v'}\sum_{v''}{|\braket{\phi_{v'}^e|\phi_{v''}^g}|^2}.
\label{eq:CMrate}
\end{equation}
Note that Eq. (\ref{eq:CMrate}) should be employed with care in the case of PA spectra close to the $^{87}$Rb$(5s\,^2S)$+$^{84}$Sr($5s5p\,^3P_{0,1,2}$) asymptotes as the TDM vanishes at large distances. It is actually more interesting to focus on the vibrational distribution in the ground state levels after RE, which is more sensitive to the $R$-variation of the electronic TDMs than the total UMF rate. We express the probability $P(v''\leftarrow v')$ of a ground-state level $v''$ to be occupied after RE from the photoassociated level $v'$ as 
\begin{equation}
P(v''\leftarrow v')=\frac{|\braket{\phi_{v'}^e|\mu_{eg}|\phi_{v''}^g}|^2}{\sum_{v''}{|\braket{\phi_{v'}^e|\mu_{eg}|\phi_{v''}^g}|^2}}.
\label{eq:vibdist}
\end{equation}

\subsection{The $^{87}$Rb$(5s\,^2S)$+$^{84}$Sr($5s5p\,^3P_{0,1,2}$) dissociation limit}

\begin{figure}
		\centering
		\includegraphics[width=10cm]{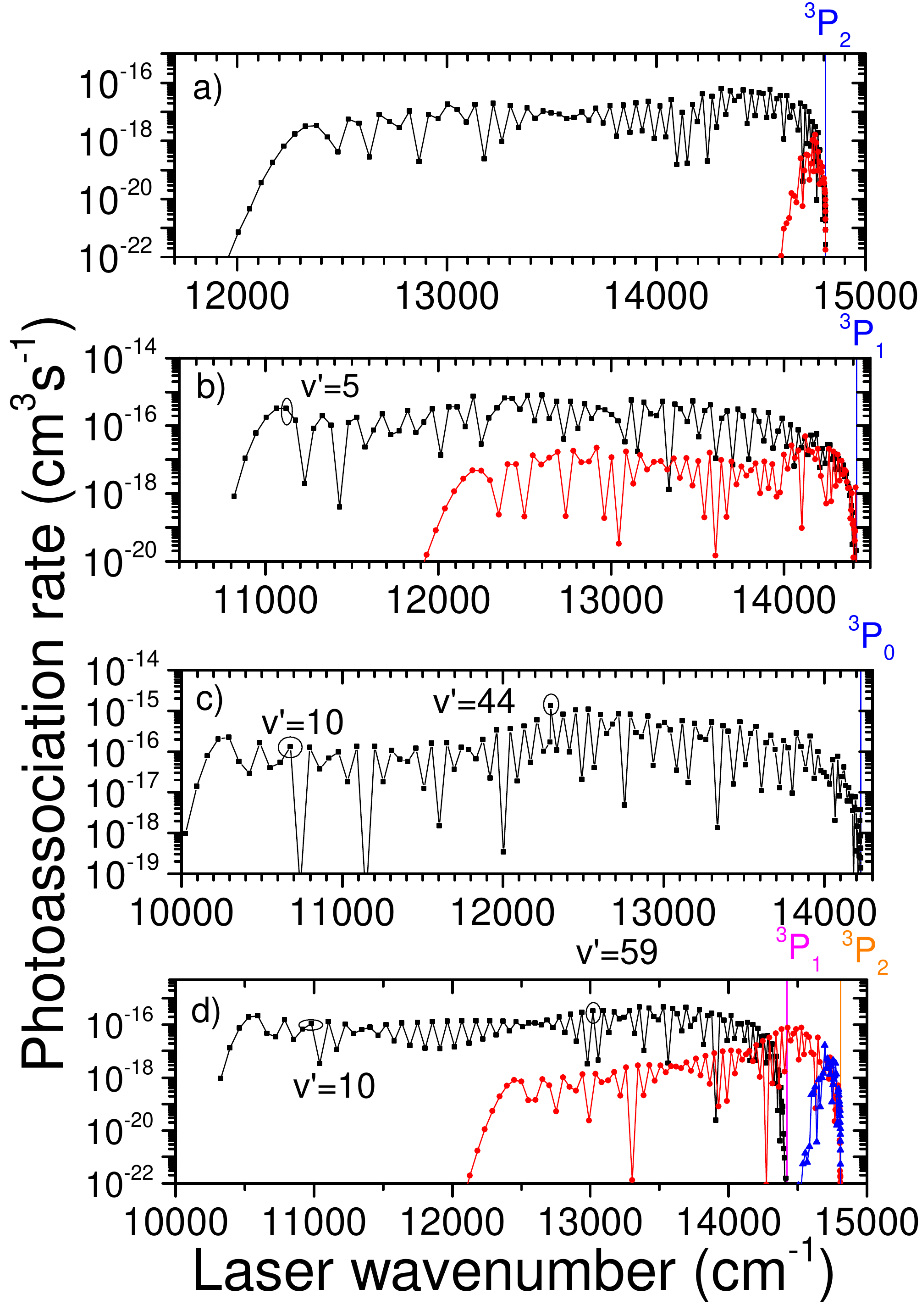}
		\caption{Photoassociation rates for $^{87}$Rb$^{84}$Sr levels as a function of the PA laser wavenumber, for states correlated to the $^{87}$Rb$(5s\,^2S)$+$^{84}$Sr($5s5p\,^3P_{0,1,2}$) dissociation limit. (a) (7) $\frac{1}{2}$ in black, and (8) $\frac{1}{2}$ in red. (b) (5) $\frac{1}{2}$ in black, and (6) $\frac{1}{2}$ in red. (c) (4) $\frac{1}{2}$ in black. (d) (2) $\frac{3}{2}$ in black, (3) $\frac{3}{2}$ in red, (4) $\frac{3}{2}$ in blue.}
		\label{fig:forbiddenPA}
	\end{figure}

PA rates for the five $\Omega=\frac{1}{2}$ states are shown on Fig. \ref{fig:forbiddenPA}. For each potential, the analysis of these PA spectra could be divided in three parts : deeply levels, intermediate levels and levels close to the asymptotes. For the deeply levels, the PA rate depends on their main composition of Hund case (a) states. The most important values are obtained for (4) $\frac{1}{2}$ and (5) $\frac{1}{2}$ that are composed by respectively (2) $^2\Pi$ and (3) $^2\Sigma$ states. Some of these deeply levels of (4) $\frac{1}{2}$ and (5) $\frac{1}{2}$ have a PA rates with only one order of magnitude lesser than the largest ones. These high values can be explained by the relative position of the (4) $\frac{1}{2}$ and (5) $\frac{1}{2}$ potential wells  with respect to the ground state one that favors the contribution  of the distances at the inner turning point of the ground state in the PA rate.

The intermediate levels have the largest PA rates. This comes from the competition between the FC factors and the value of TDMs at the outer turning point as illustrated on Fig.\ref{fig:squaredTDME}. Due to their most important composition in (2) $^2\Pi$ and (3) $^2\Sigma$ states, the largest rates are obtained with the states (4) $\frac{1}{2}$ and (5) $\frac{1}{2}$ : $1.37\times 10^{-15}$ ~cm$^{-3}$s$^{-1}$ at 12295.1~cm$^{-1}$ ($v'=44$), and $5.37\times 10^{-16}$  ~cm$^{-3}$s$^{-1}$ at 12601.3~cm$^{-1}$ ($v'=36$), respectively. For the (6),(7) and (8) $\Omega=\frac{1}{2}$ states, the PA rates significantly increase, due to the admixture of doublet states in addition to the quartet component at large interatomic separation, but remain smaller. 

Finally, close to the asymptotes, the PA rates significantly drop down for all states, for example at a detuning around 0.15~cm$^{-1}$, the PA rates take the values : $7.2\times 10^{-20}$~cm$^{-3}$s$^{-1}$ and $2.7\times 10^{-22}$ ~cm$^{-3}$s$^{-1}$ for respectively the (4) $\frac{1}{2}$ and (5) $\frac{1}{2}$ states. Therefore PA already appears as a challenge close to the asymptote. New results for PA in RbSr are available \cite{ciamei2018}. In particular, the authors report difficulties to observe PA close to the $^{87}$Rb$(5s\,^2S)$+$^{84,87}$Sr($5s5p\,^3P_{1}$) asymptote, and measure small photoassociation rates. Such low rates are consistent with the low rates of our calculations.  

The results for the $\frac{3}{2}$ are rather similar (Fig. \ref{fig:forbiddenPA}d). The maximal PA rate ($4.7\times 10^{-16}$~cm$^{-3}$s$^{-1}$ for a detuning of 13018.32~cm$^{-1}$ ($v'=59$)) is found for the (2) $\frac{3}{2}$, mainly composed of doublet states. Close to the asymptote, they also become rather low, remaining for the (4) $\frac{3}{2}$ as large as the $\frac{1}{2}$ states.

\begin{figure}
		\centering
		\includegraphics[width=12cm]{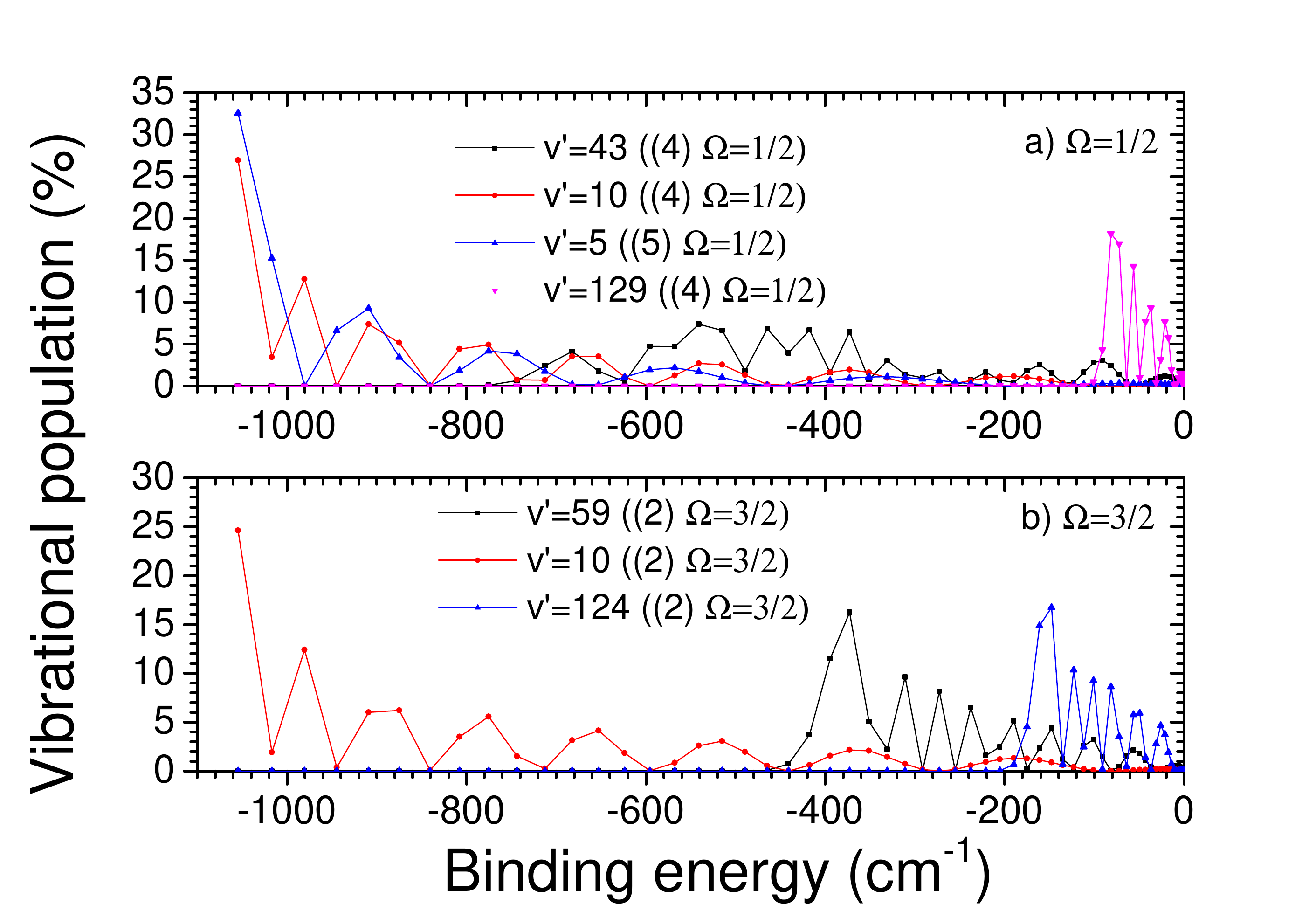}
		\caption{Distribution of vibrational populations (displayed in percentage) of the ultracold $^{87}$Rb$^{84}$Sr molecules created after spontaneous emission from several photoassociated levels belonging to states converging toward the $^3P_0$ ((4) $\frac{1}{2}$ ) and $^3P_1$ ((5)$\frac{1}{2}$ and (2) $\frac{3}{2}$ ) limits . (a) the $v'=43$ (black line), $v'=10$ (red line)  and $v'=129$ (pink line)  levels of (4)$\frac{1}{2}$ and the $v'=5$ (blue line) level of (5)$\frac{1}{2}$. (b) the $v'=59$ (black line), $v'=11$ (red line) and $v'=124$ (blue line) levels of (2) $\frac{3}{2}$.}
		\label{fig:forbiddenvibdist}
\end{figure}

The computed energy variations of the UMF rates are found very similar to those of the PA rates, and are displayed in the Appendix for the sake of conciseness. Using Eq. (\ref{eq:CMrate}), we find that  the sum of squared overlap in Eq. (\ref{eq:CMrate}) decreases for (4) $\frac{1}{2}$ and (5) $\frac{1}{2}$ while increases for (6) $\frac{1}{2}$ and (7) $\frac{1}{2}$ due to the admixtures between doublet and quartet states as already explained for the PA rates. However, over the entire range, it is always in the same order of magnitude and does not induce a change of the spectra of UMF rates with respect to the PA one. The vibrational distributions in the ground-state levels (see Eq. (\ref{eq:vibdist})) are displayed in Fig. \ref{fig:forbiddenvibdist} for few typical photoassociated levels $v'$ : deeply levels ($v'=5$ of (4) $\frac{1}{2}$, $v'=10$ of (5) $\frac{1}{2}$ and $v'=11$ of (2)$\frac{3}{2}$), intermediate levels ($v'=43$ of (4) $\frac{1}{2}$ and $v'=59$ of (2) $\frac{3}{2}$) that have the largest PA rates, and the last bound levels ($v'=129$ of (4) $\frac{1}{2}$ and $v'=124$ of (2) $\frac{3}{2}$). The deeply-bound levels can yield a main fraction of the population in the ground state $v''=0$ level, as it can be expected from the favorable relative position of the minimum of the corresponding potential well in the excited and in the ground state. The levels with the largest PA rates (in the (4)$\frac{1}{2}$ and the (2)$\frac{3}{2}$ states) induce a population spread over numerous ground state vibrational levels, with no specific emergence of a particular level (a maximum population being found however for respectively level $v''$=16 ($E_{bind}$=-541~cm$^{-1}$) and $v''$=23 ($E_{bind}$=-372~cm$^{-1}$)). Finally, for the last bound levels, different levels with binding energies below 200~cm$^{-1}$ are populated. The population goes to zero for the last bound levels of ground state and it is a consequence of the vanishing TDM at large distance.

\subsection{The Rb$(5p\,^2P_{1/2,3/2})$+Sr($5s^2\,^1S$) dissociation limit}
\label{ssec:allowed}

\begin{figure}
		\centering
		\includegraphics[width=12cm]{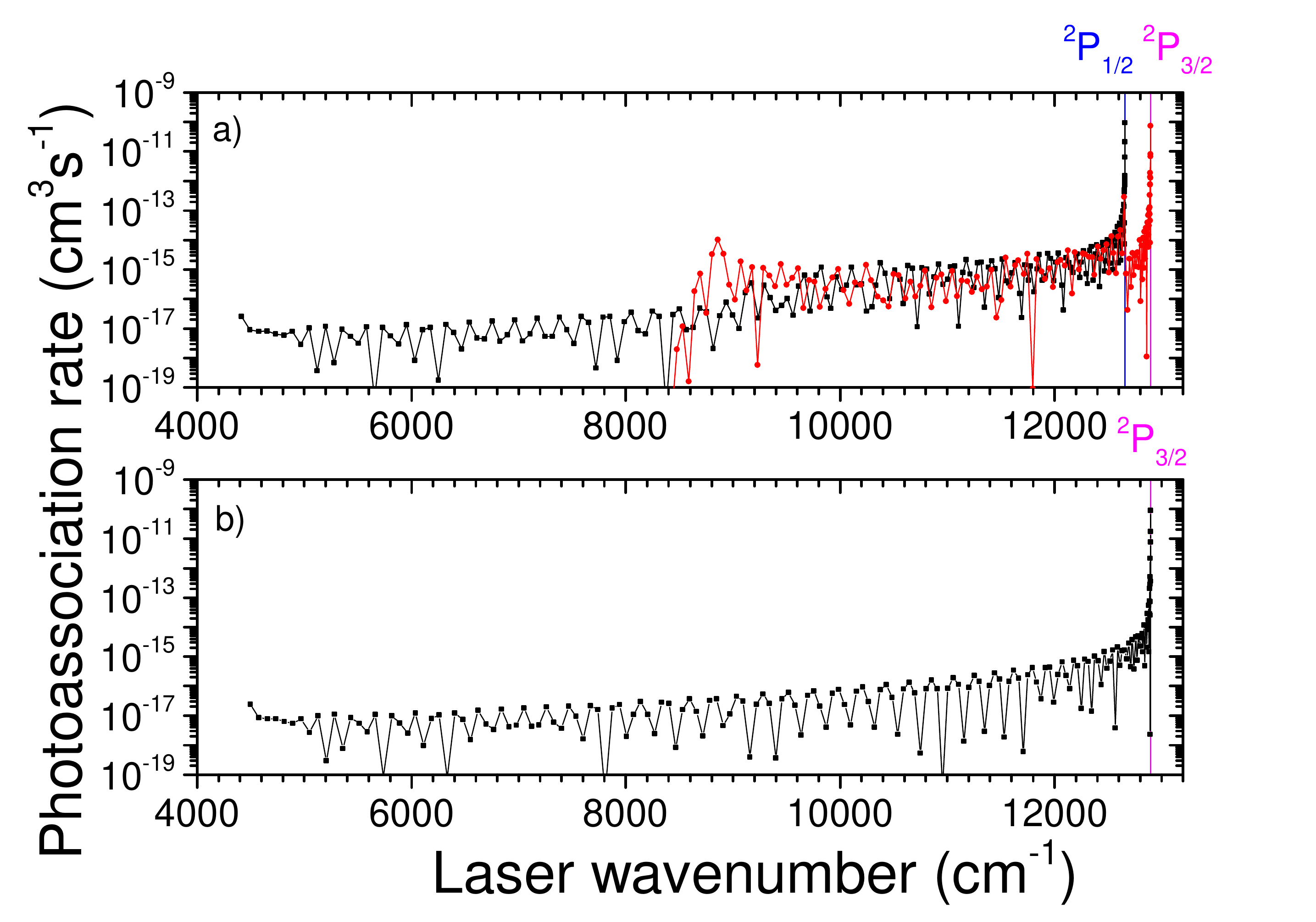}
		\caption{Photoassociation rates for the bound levels of (a) the (2) $\frac{1}{2}$, (3)$\frac{1}{2}$ states, and (b) of the (1)$\frac{3}{2}$ state, correlated to the Rb$(5p\,^2P_{1/2,3/2})$+Sr($5s^2\,^1S$) dissociation limit and identified by the laser wavenumber.}
		\label{fig:allowedPA}
\end{figure}

The computed PA rates for both $\Omega=\frac{1}{2}$ and $\Omega=\frac{3}{2}$ states are displayed in Fig. \ref{fig:allowedPA}. The analysis of these PA spectra could be also divided for deeply-bound levels on one hand, and for levels close to the asymptotes on the other hands. For deeply levels, the PA rates are more important for (2) $\Omega=\frac{1}{2}$ than for (3) $\Omega=\frac{1}{2}$. This striking difference is at first sight counter-intuitive. The levels from (2) $\frac{1}{2}$ state composed mainly of $^2\Pi$ (low value at short-range) state have larger PA rates than levels from (3) $\frac{1}{2}$ state composed mainly of $^2\Sigma^+$ (high value at short-range). A deeper study show that the main parameter is the overlaps and so the FCFs that are specially weak for the deeply-bound levels of (3) $\frac{1}{2}$ state. Close to the asymptotes, PA rates quickly rise. The results are very similar to those obtained for the alkali-metal dimers, as the dominant role is given to the dipole-allowed Rb transition. In a classical view, PA takes place mostly at large interatomic distances, where the electronic TDM is large, and where the overlap between the relevant radial wave functions is favored. This could be seen on Fig.\ref{fig:squaredTDME} b). The rate magnitude is found similar to the one for Rb photoassociation \cite{Rb2_Fioretti}. The maximal values ($10^{-10}-10^{-11}$~cm$^{-3}$s$^{-1}$) close to the dissociation limits are tedious to compare to experiment, as the cloud of cold atoms is strongly perturbed if the PA laser is tuned too close to the atomic resonance. 

\begin{figure}
		\centering
		\includegraphics[width=8cm]{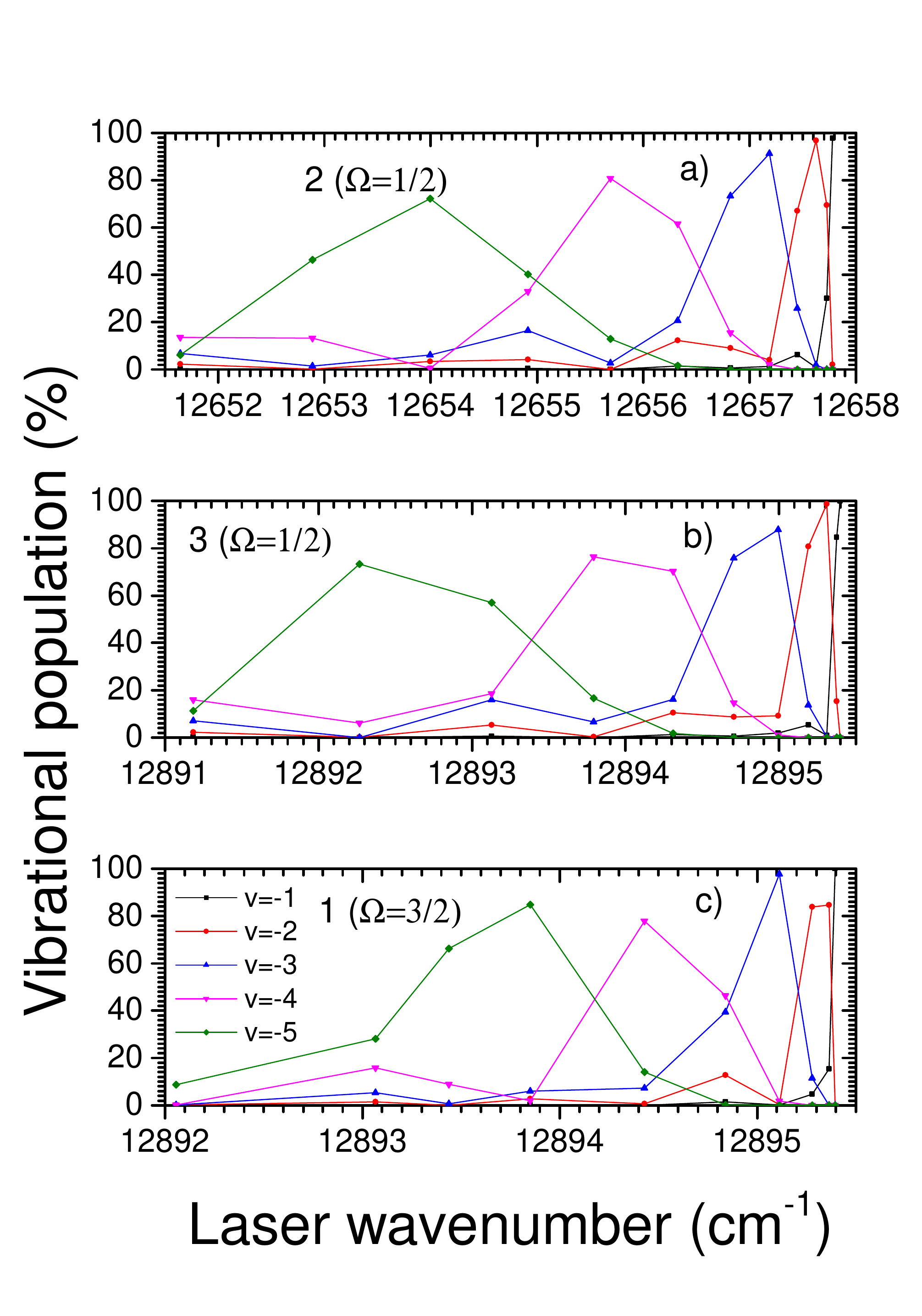}
		\caption{Distribution of vibrational populations (displayed in percentage) of the five uppermost vibrational levels (labeled as $v=-1, -2, -3, -4, -5$ for convenience) of the $^{87}$Rb$^{84}Sr$ ground state, generated after spontaneous emission from the photoassociated levels of the (2) $\frac{1}{2}$, (3) $\frac{1}{2}$, and (1) $\frac{3}{2}$ states (panels (a), (b) and (c), respectively) identified by the PA laser wavenumber. }
		\label{fig:allowedvibdist}
\end{figure}

As previously, the UCM rate variations with the PA laser frequency are very similar to the ones of the PA rates, and are displayed in the Appendix. The computed vibrational distributions generated by the photoassociated levels are actually quite remarkable, as illustrated in Fig.\ref{fig:allowedvibdist} for the uppermost PA levels. They reflect an almost diagonal Franck-Condon matrix, namely each photoassociated level almost populates a single ground state level: for instance, the last-but-one ground-state level (noted $v''=-1$ for convenience) is populated at 97.75\% by the last-but one (also noted $v'=-1$ for convenience) level of (2) $\frac{1}{2}$, at 99.98 \% with the $v'=-1$ level of (3) $\frac{1}{2}$, and at 99.94\% of (1) $\frac{3}{2}$. The purity of the relaxation process is worse when the detuning is increased. But in any case, only weakly-bound ground state molecules could be efficiently created.  Note that this situation has been also investigated in Ref. \cite{Chen_RbSr} based only on the Franck-Condon factors, yielding results similar to ours.

\section{Formation of ultracold weakly-bound molecules by a STIRAP transfer in a tight optical trap}
\label{sec:STIRAPtrap}

In the previous section, we have shown that the formation of RbSr ground-state molecules could be achieved by PA. However, the spontaneous emission induces two well-known drawbacks: the loss of atoms from the trap, and a broad distribution of occupied vibrational levels. Even if this last point is minimized when PA is implemented with a laser frequency close to the dipole-allowed Rb transition (see Section \ref{ssec:allowed}), employing a coherent method avoiding spontaneous emission would decrease the loss of atoms. The most efficient method of coherent population transfer is the STimulated Rapid Adiabatic Passage (STIRAP) method \cite{STIRAP_1} relying on a proper choice of three energy levels, refereed to as a $\Lambda$ system. It allows the coherent population transfer between an initial level $\ket{i}$ and a final level $\ket{f}$ through a dark state involving an excited level $\ket{e}$. When the two-photon detuning $\delta$ is zero, one of the instantaneous eigenstates of the Hamiltonian including the light is a coherent superposition of only $\ket{i}$ and $\ket{f}$ and is the so-called dark state. The transfer could be made without populating the intermediate level $\ket{e}$, thereby avoiding loss by spontaneous emission. The complete population transfer from $\ket{i}$ to $\ket{f}$ is achieved by the application of a pair of pump and dump pulses in a counter-intuitive order. As the system must remain in the dark state during all the process, the dynamics has to be adiabatic, thus imposing constraints on the corresponding Rabi frequencies $\Omega_{\textrm{pump}}$ and $\Omega_{\textrm{dump}}$: they must have the same order of magnitude, and they must be sufficiently large to satisfy the condition $\Omega_{\textrm{pump,dump}} T>>\pi$ where $T$ is the duration of the pulses.

The use of STIRAP in a PA experiment (\textit{i.e.} without relying on Feshbach resonances) was previously investigated \cite{STIRAP_photoassociation_1}, revealing some difficulties due to the fact that the initial level belongs to a dissociation continuum, preventing the perfect creation of the dark state. One can overcome this drawback by placing the initial cold atoms in a tight optical trap (say, at a typical wavelength of 1064~nm), such that the motional states of the atom pair become quantized \cite{jaksch2002}. At ultracold temperature, the atoms occupy the lowest motional level of the trap. Therefore, the radial wave function of the atom pair should be  localized at shorter distance, and the Franck-Condon factors with the bound levels of excited electronic states should increase. 

The Hamiltonian describing two non-identical atoms of mass $m_1$ and $m_2$ at positions $\vec{r}_1$ and $\vec{r}_2$ in an optical anharmonic trap with harmonic frequencies $\omega_1$ and $\omega_2$ felt by each atomic species, is \cite{Tight_trap_1}
\begin{equation}
H_{\textrm{trap}}=-\frac{\hbar^2}{2M}\Delta_{\textrm{com}}+\frac{1}{2} M \omega^2_{\textrm{com}} R_{\textrm{com}}^2-\frac{\hbar^2}{2\mu}\Delta_{R}+\frac{1}{2}\mu\omega^2_{R}R^2+V(R)+\mu\Delta\omega\vec{R}_{\textrm{com}}.\vec{R}+V_{\textrm{anharm}}
\label{eq:Htrap}
\end{equation}
with the total mass $M=m_1+m_2$, the reduced mass $\mu=\frac{m_1m_2}{M}$, the position of the center-of-mass $\vec{R}_{\textrm{com}}=\frac{m_1\vec{r}_1+m_2\vec{r}_2}{M}$, the relative position vector $\vec{R}=\vec{r}_1-\vec{r}_2$, and $\Delta\omega=\sqrt{\omega_1^2-\omega_2^2}$. The first two terms represent the center-of-mass motion in the trap, with frequency $\omega_{\textrm{com}}=\sqrt{\frac{m_1\omega_1^2+m_2\omega_2^2}{m_1+m_2}}$. The next three terms describe the relative motion of the atom pair interacting through the potential $V(R)$, in the presence of a trapping potential of frequency $\omega_{R}=\sqrt{\frac{m_2\omega_1^2+m_1\omega_2^2}{m_1+m_2}}$. These two motions are in principle coupled by the anharmonic terms $V_{\textrm{anharm}}$ of the trapping potential, and by a dynamical term proportional to $\vec{R_{\textrm{com}}}.\vec{R}$. The former can be safely neglected if we assume that the atoms are trapped in the lowest motional level. The latter depends on the differences of masses and polarizabilities that are almost the same in our case. In our calculation, we have therefore neglected the coupling between the two motions, and worked with relative coordinate. We have taken the experimental trapping frequencies $2\pi\times 65$~kHz for $^{84}$Sr and $2\pi \times 110$~kHz for $^{87}$Rb \cite{Schreck_pc}. The characteristic length of the relative motion in the trap is $a_{\omega}=\sqrt{\hbar/\mu\omega_{rel}}=969$ a.u. which is much larger than the scattering length. Therefore, the tight trap does not induce any significant modification of bound levels of the ground and excited molecular states. The eigenstates of $H_{\textrm{trap}}$ for the ground and excited states of Fig.\ref{fig:PEC_SO} as well as the transition matrix elements are computed with the same procedure than in the previous sections. The main difference is that the radial wave functions of the trap states are now normalized to unity, as they are no longer continuum states. 

The initial level $\ket{i}$ is taken as the first trap state. For the final level $\ket{f}$, we have first chosen the vibronic ground-state level ($v''=0$). In addition, we have examined the possibility to improve the STIRAP process toward another final level. The crucial element of the model is the choice of the best possible intermediate level $\ket{e}$ belonging to an excited electronic state. Two requirements  must be considered, independently of the experimental laser intensities used in the experiment: the squared matrix elements of the transition dipole moment (squared TDMEs in short) for the pump and dump transitions must be of the same order of magnitude, and be sufficiently high (typically more than $10^{-6}$ a.u., see for instance Ref.\cite{borsalino2014}). 

\begin{figure}
		\centering
		\includegraphics[width=8cm]{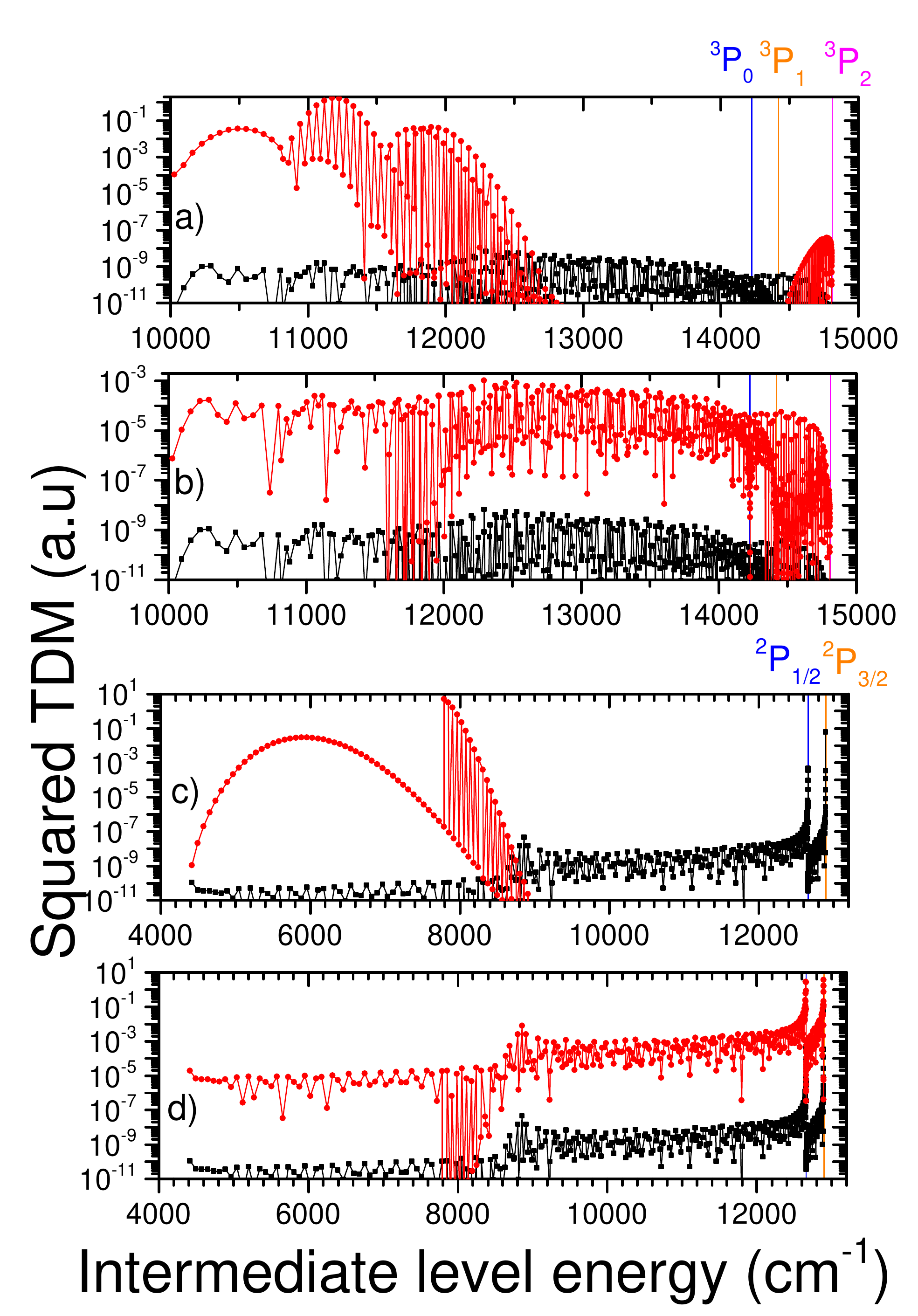}
		\caption{Squared matrix elements of the transition dipole moment (squared TDMEs in short) relevant for the formation of $^{87}$Rb$^{84}$Sr molecules with STIRAP, starting from an atom pair confined in the lowest motional level of a tight optical trap (see text for details), as a function of the excitation energy of the chosen intermediate level $\ket{e}$. Pump transitions: black lines; dump transitions: red lines. Panels (a) and (c) (rep. (b) and (d)) correspond to the final level $\ket{f} \equiv v''=0$ (resp. $\ket{f} \equiv v''=-3$) of the electronic ground state. The levels $\ket{e}$ belong to all electronic states $\Omega=\frac{1}{2}$ correlated to $^{87}$Rb$(5s\,^2S)$+$^{84}$Sr($5s5p\,^3P_{0,1,2}$) (panels (a) and (b)), and  to Rb$(5p\,^2P_{1/2})$+Sr($5s^2\,^1S$)) (panels (c) and (d)).}
		\label{fig:STIRAPtrap}
\end{figure}

For the transfer toward $\ket{f} \equiv \ket{v''=0}$ via the states $\ket{e}$ correlated to $^{87}$Rb$(5s\,^2S)$+$^{84}$Sr($5s5p\,^3P_{0,1,2}$) (Fig. \ref{fig:STIRAPtrap}a), the squared TDMEs curves for the pump and dump transitions cross twice each other, around 12000~cm$^{-1}$ and 14500~cm$^{-1}$ with very weak magnitudes ($10^{-9}-10^{-10}$~a.u.). A similar conclusion holds for the transfer via the levels $\ket{e}$ belonging to states correlated to Rb$(5p\,^2P_{1/2,3/2})$+Sr($5s^2\,^1S$), where the two curves cross once around 8500~cm$^{-1}$ (Fig. \ref{fig:STIRAPtrap}c) with a low magnitude ($10^{-9}$~a.u.). These statements actually reflect the behavior of the corresponding PA rates of Figs. \ref{fig:forbiddenPA} and \ref{fig:allowedPA}. While providing a discrete level for the initial state for STIRAP, the choice of a confined trap level for the pump step does not significantly improve the magnitude of the squared TDMEs compared to the conventional PA starting from a real continuum state.

Panels (b) and (d) in Fig. \ref{fig:STIRAPtrap} illustrate another possible way to progress on the way to the creation of ultracold RbSr molecules. We have calculated the TDMEs involved in the transfer from the initial trap state toward the $v=-3$ level of the RbSr ground state, leading to a contrasted result: while the STIRAP transfer does not seem to be possible via $\ket{e}$ levels belonging to states correlated to $^{87}$Rb$(5s\,^2S)$+$^{84}$Sr($5s5p\,^3P_{0,1,2}$) (Fig. \ref{fig:STIRAPtrap}b)), it appears doable via $\ket{e}$ levels close to the Rb$(5p\,^2P_{1/2,3/2})$+Sr($5s^2\,^1S$) dissociation limit (see the extreme right part of the Fig. \ref{fig:STIRAPtrap} d), for which the squared TDMEs for the pump and dump transitions can reach similar values up to 10$^{-5}$~a.u. In fact, the last five bound levels could be populated by a STIRAP with an intermediate level close to the asymptotes $^2P_{1/2}$ or $^2P_{3/2}$ (see table \ref{STIRAP_trap_allo_trans}). As expected, the STIRAP is more tedious for deeper final levels. In conclusion, the STIRAP method in a tight trap can create only weakly-bound $^{87}$Rb$^{84}$Sr ground state molecules, just like PA or MFR.

\begin{table}
\centering
\caption{Characteristics of the optimal transitions for the STIRAP scheme in a tight trap via an intermediate level close to Rb$(5p\,^2P_{1/2})$+Sr($5s^2\,^1S$)). The initial level is the first trap state, with an energy of $5.10^{-6}$~cm$^{-1}$ (or about 150~kHz) above Rb$(5s\,^2S_{1/2})$+Sr($5s^2\,^1S$). The final level of the ground state is labeled with negative index $\tilde{v}_f$ starting from the Rb$(5s\,^2S_{1/2})$+Sr($5s^2\,^1S$) asymptote, with a binding energy $E_f$. The vibrational index $v_e$ and binding energy $E_{e}$ of several optimal intermediate levels are displayed. The energies $E_{\textrm{pump}}$ and $E_{\textrm{dump}}$, and the related squared transition dipole moments $|d_{ie}|^2$ and $|d_{ef}|^2$ of the pump and dump transitions are also reported. Numbers in parenthesis hold for powers of 10.}
\begin{tabular}{cccccc}
\hline
\hline
$\tilde{v}_f$      & -1             & -2             & -3             & -4             & -5 \\
$E_{f}$ ($cm^{-1}$)& 1.3(-3)        & 2.58(-2)       & 1.147(-1)      & 3.112 (-1)     & 6.573 (-1) \\\hline
$v_e$              & 161            & 201            & 199            & 198            & 197  \\
                   &3($\frac{1}{2}$)&2($\frac{1}{2}$)&2($\frac{1}{2}$)&2($\frac{1}{2}$)&2($\frac{1}{2}$) \\
$E_{e}$ ($cm^{-1}$)& 2.9 (-3)       & 2.0(-3)        & 7.16(-2)       & 1.755 (-1)     & 3.494 (-1) \\\hline
$E_{\textrm{pump}}$ (cm$^{-1}$)&12895.4000&12657.7979& 12657.7279     & 12657.6238     & 12657.4490 \\
$|d_{ie}|^2$ (a.u.)             & 6.2 (-2)  & 4.5 (-4)  & 9.3 (-5)       & 2.8 (-6)       & 2.8 (-5) \\
$E_{\textrm{dump}}$ (cm$^{-1}$)&12895.4012&12657.8238& 12657.8426     & 12657.9350     & 12658.1063 \\
$|d_{ef}|^2$ (a.u.)             & 1.2 (0)  & 7.4 (-4) & 2.3 (-4)       & 7.1 (-6)       & 2.5 (-5) \\
\hline
\hline
\end{tabular}
\label{STIRAP_trap_allo_trans}
\end{table}

\section{Population transfer from weakly-bound RbSr ground-state molecules to the rovibrationnal ground state}
\label{sec:STIRAPgroundstate}

In the two last sections, we have shown that the formation of weakly-bound $^{87}$Rb$^{84}$Sr ground-state molecules is achievable by PA and by STIRAP in a tight trap. A second STIRAP step could then be implemented to transfer these molecules into the lowest level of the ground state. Such a double STIRAP sequence has already been applied for ultracold Cs$_2$ molecules \cite{danzl2010}. We have looked for an optimal STIRAP transfer starting from the five uppermost ground state levels above, now labeled as $v_i=-1, -2, -3, -4, -5$.

\begin{figure}
		\centering
		\includegraphics[width=12cm]{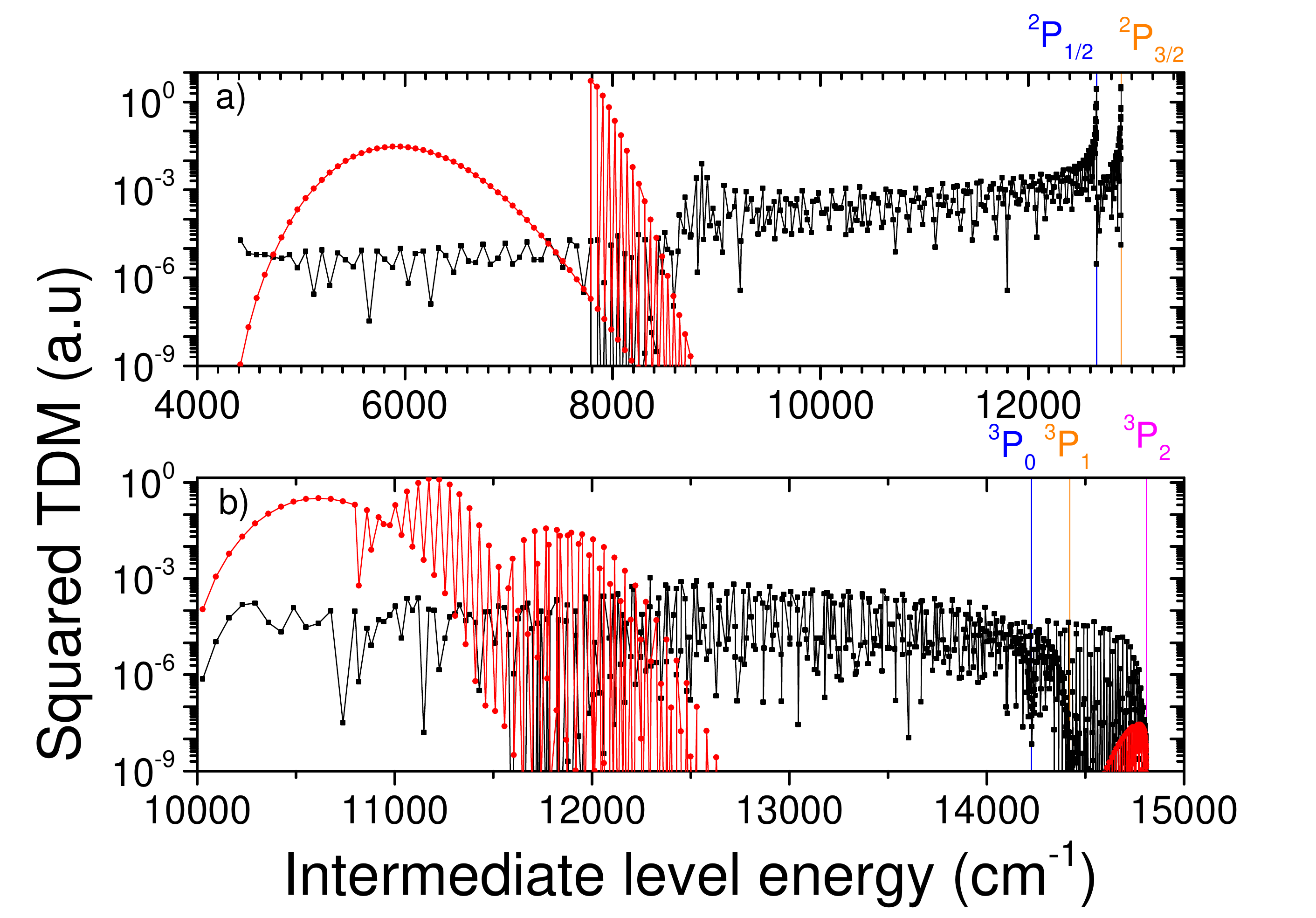}
		\caption{Squared matrix elements of the transition dipole moment (squared TDMEs in short) relevant for transferring population from v''=-3 to v''=0  with STIRAP as a function of the excitation energy of the chosen intermediate level $\ket{e}$. Pump transitions: black lines; dump transitions: red lines. Panels (a) (resp. (b)) correspond to STIRAP with intermediate levels belonging to all electronic states correlated to Rb$(5p\,^2P_{1/2})$+Sr($5s^2\,^1S$) (resp. $^{87}$Rb$(5s\,^2S)$+$^{84}$Sr($5s5p\,^3P_{0,1,2}$}
		\label{fig:STIRAPweak}
	\end{figure}

We have identified three efficient STIRAP paths in three different spectral zones, based upon the same criterion than above of the equality of the squared TDMs for the pump and dump transitions:
\begin{itemize}
	
	\item the first scheme relies on intermediate levels of the 2($\frac{1}{2}$) state correlated to Rb$(5p\,^2P_{1/2})$+Sr($5s^2\,^1S$)), corresponding to $E_{\textrm{pump}}$ in the 4570-4890~cm$^{-1}$ range, and $E_{\textrm{dump}}$ in the 5625-5945~cm$^{-1}$ range (Table \ref{STIRAP_weakly_schema_1}, and Fig\ref{fig:STIRAPweak}a). Despite a strong magnitude of the corresponding squared TDMs, that may not be the most practical frequencies to implement experimentally. As already noticed before, the possibility to use the lowest bound levels of the intermediate state comes from the relative position of PECs, and from the position of the inner turning point of the initial weakly-bound vibrational wave function, located close to the equilibrium distance of excited states. 
	
	\item The second scheme relies on the same intermediate state, with levels that can be reached with $E_{\textrm{pump}}$ in the 7175-7590~cm$^{-1}$ range, inducing $E_{\textrm{dump}}$ located in the 8230-8640~cm$^{-1}$ range (Table \ref{STIRAP_weakly_schema_2}, and Fig\ref{fig:STIRAPweak}a). This scheme is expected to be slightly less efficient than the previous one, but in a more accessible frequency domain for the STIRAP lasers. This solution involves the vibrational levels close to the avoided crossing between the $^2\Sigma^+$ and $^2\Pi$ states (see fig. \ref{fig:STIRAPtrap} ).
	
	\item The third scheme involves levels of the $4 (\frac{1}{2})$ state correlated to Rb$(5s\,^2S)$+Sr($5s5p\,^3P_2$) (Table \ref{STIRAP_weakly_schema_3}, and Fig\ref{fig:STIRAPweak}b), with $E_{\textrm{pump}}$ in the 11200-11360~cm$^{-1}$ range, and $E_{\textrm{dump}}$ in the 12255-12415~cm$^{-1}$ range. This corresponds to levels with an energy close to the avoided crossing visible in Fig. \ref{fig:STIRAPtrap}. The efficiency of this STIRAP path seems to be the best of the three presented in this work. The advantage of this path is that a laser Ti:sapphire could be used. 
\end{itemize} 

Chen \textit{et al.} \cite{Chen_RbSr} have proposed another STIRAP path using the $v'=21$ level of the (2) $\frac{1}{2}$ state as the intermediate level, relying on a hypothesis different than ours:  the selected intermediate level should be the one with the largest value for the product of the squared TDMs for the pump and dump transitions (actually reduced to FCF in their paper). The drawback of such a methodology is that the squared TDMs for the pump and the dump transitions could be vastly different, thus implying very different laser intensities. It is indeed the case here, as there are 4 orders of magnitude of difference for squared TDM. The advantage of STIRAP path presented in our work is that the intensities for the two transitions would be similar.  

\begin{table}
\centering
\caption{Characteristics of the optimal transition for the first STIRAP scheme with an intermediate level belonging to an PEC correlated to the asymptotes $^2P_{1/2}$ and $^2P_{3/2}$.The final level is the rovibrational ground state ($v''=0$). The vibrational number and binding energy $E_{e}$ are given for the intermediate level. The energies $E_{\textrm{pump}}$ and $E_{\textrm{dump}}$ and the related squared transition dipole moments $|d_{ie}|^2$ and $|d_{ef}|^2$ of the pump and stoke transition are also reported. Number in parenthesis hold for powers of 10.}
\begin{tabular}{cccccc}
\hline
\hline
$v_i$ & -1 & -2 & -3 & -4 & -5 \\
$E_{i}$ (cm$^{-1}$) & 1.3(-3) & 2.58(-2) & 1.147(-1) & 3.112 (-1) & 6.573 (-1) \\
\hline
$v_f$ & 0 & 0 & 0 & 0 & 0 \\
$E_{f}$ (cm$^{-1}$) & 1054.3406 & 1054.3406 & 1054.3406 & 1054.3406 & 1054.3406 \\
\hline
$v_e$ ((2) $\Omega=\frac{1}{2}$) & 4 & 4 & 5 & 6 & 6 \\
$E_{e}$ (cm$^{-1}$) & 7926.4524 & 7926.4524 & 7847.7908 & 7769.4210 & 7769.4210 \\
\hline
$E_{\textrm{pump}}$ (cm$^{-1}$) & 4731.3473 & 4731.3473 & 4810.3193 & 4889.0356 & 4889.0356 \\
$|d_{ie}|^2$ (a.u.) & 2.1 (-7) & 1.7 (-6) & 4.5 (-6) & 1.2 (-5) & 2.0 (-5) \\
$E_{\textrm{dump}}$ (cm$^{-1}$) & 5785.6882 & 5785.6882 & 5864.3497 & 5942.7196 & 5942.7196 \\
$|d_{ef}|^2$ (a.u.) & 6.1 (-6) & 6.1 (-6) & 2.4 (-5) & 7.6 (-5) & 7.6 (-5) \\
\hline
\hline
\end{tabular}
\label{STIRAP_weakly_schema_1}
\end{table}

\begin{table}
\centering
\caption{Same as Table \ref{STIRAP_weakly_schema_1} for the second STIRAP scheme with an intermediate level belonging to a PEC correlated to the asymptotes $^2P_{1/2}$ and $^2P_{3/2}$.}
\begin{tabular}{cccccc}
\hline
\hline
$v_i$ & -1 & -2 & -3 & -4 & -5 \\
$E_{i}$ (cm$^{-1}$) & 1.3(-3) & 2.58(-2) & 1.147(-1) & 3.112 (-1) & 6.573 (-1) \\
\hline
$v_f$ & 0 & 0 & 0 & 0 & 0 \\
$E_{f}$ (cm$^{-1}$) & 1054.3406 & 1054.3406 & 1054.3406 & 1054.3406 & 1054.3406 \\
\hline
$v_e$ ((2) $\Omega=\frac{1}{2}$) & 40 & 40 & 39 & 37 & 37 \\
$E_{e}$ (cm$^{-1}$) & 5275.7563 & 5275.7563 & 5344.3918 & 5482.5236 & 5482.5236 \\
\hline
$E_{\textrm{pump}}$ (cm$^{-1}$) & 7382.0436 & 7382.0687 & 7313.5220 & 7175.5865 & 7175.9330 \\
$|d_{ie}|^2$ (a.u.) & 2.5(-7) & 2.1 (-6) & 1.4 (-6) & 1.1 (-5) & 1.9 (-5) \\
$E_{\textrm{dump}}$ (cm$^{-1}$) & 8436.3843 & 8436.3843 & 8367.7488 & 8229.6170 & 8229.6170 \\
$|d_{ef}|^2$ (a.u.) & 1.8 (-6) & 1.8 (-6) & 3.8 (-6) & 1.6 (-5) & 1.6 (-5) \\
\hline
\hline
\end{tabular}
\label{STIRAP_weakly_schema_2}
\end{table}

\begin{table}
\centering
\caption{Same as Table \ref{STIRAP_weakly_schema_1} for the third STIRAP scheme with an intermediate level belonging to a PEC correlated to the asymptotes $^3P_{0}$,$^3P_{1}$ and $^3P_{2}$.}
\begin{tabular}{cccccc}
\hline
\hline
$v_i$ & -1 & -2 & -3 & -4 & -5 \\
$E_{i}$ (cm$^{-1}$) & 1.3(-3) & 2.58(-2) & 1.147(-1) & 3.112 (-1) & 6.573 (-1) \\
\hline
$v_f$ & 0 & 0 & 0 & 0 & 0 \\
$E_{f}$ (cm$^{-1}$) & 1054.3406 & 1054.3406 & 1054.3406 & 1054.3406 & 1054.3406 \\
\hline
$v_e$ ((5) $\Omega=\frac{1}{2}$) & 16 & 15 & 15 & 15 & 15 \\
$E_{e}$ (cm$^{-1}$) & 2746.0037 & 2746.0131 & 2746.0131 & 2746.0131 & 2746.0131 \\
\hline
$E_{\textrm{pump}}$ (cm$^{-1}$) & 11675.2972 & 11626.3119 & 11626.4005 & 11626.5954 & 11626.9390 \\
$|d_{ie}|^2$ (a.u.) & 7.4 (-6) & 2.0 (-5) & 5.5 (-5) & 1.1 (-4) & 1.7 (-4) \\
$E_{\textrm{dump}}$ (cm$^{-1}$) & 12414.0104 & 12362.3483 & 12362.3483 & 12362.3483 & 12362.3483 \\
$|d_{ef}|^2$ (a.u.) & 1.8 (-5) & 9.8 (-5) & 9.8(-5) & 9.8 (-5) & 9.8 (-5) \\
\hline
\hline
\end{tabular}
\label{STIRAP_weakly_schema_3}
\end{table}

\section{Conclusion}
In this work, we have made a complete investigation about ways to create ultracold $^{87}$Rb$^{84}$Sr bosonic molecules in their rovibronic absolute ground state by all-optical methods. We have modeled the photoassociation of ($^{87}$Rb,$^{84}$Sr) atom pairs close to two atomic transitions: the allowed $5s^2S_{1/2} \rightarrow 5p\,^2P_{1/2,3/2}$ Rb transition, and the $5s^2\,^1S \rightarrow 5s5p\,^3P_{0,1,2}$ intercombination in strontium. As expected the photoassociation spectra show opposite behaviors. In the former case, the photoassociation rates are very high close to the asymptote. In the latter case, the photoassociation rates are very low close to the asymptotes. The distributions of ground-state vibrational levels after spontaneous emission are also different. Mainly one vibrational level is populated in the former case, but this level is highly excited. In the latter case, the lowest rovibrational level of the ground state could be populated, but many other vibrational levels as well. Therefore a further step of internal cooling is necessary to achieve a significant creation of ultracold RbSr molecules in their lowest rovibrational level.

We have then proposed to implement the formation of ultracold $^{87}$Rb$^{84}$Sr molecules by a STIRAP method in a tight trap. We found that a single STIRAP sequence to reach the lowest rovibrational ground-state level is tedious with moderate laser intensity. However, with an intermediate level close to the allowed $5s^2S_{1/2} \rightarrow 5p\,^2P_{1/2,3/2}$ Rb transition, a STIRAP schema is possible for populating one of the five last vibrational levels of the ground state. We then completed our study by modeling a further STIRAP sequence to efficiently transfer the population from these uppermost levels toward the lowest rovibrational ground-state level. Three STIRAP schemes have been identified in three different spectral zones. 

Together with the recent spectacular experimental achievements of the Amsterdam group \cite{barbe2017,ciamei2018} revealing magnetic Feshbach resonances in RbSr and a novel description of the entire PEC of the RbSr ground state, the present work should help to progress toward the realization of a molecular sample of ultracold RbSr polar molecules. From our investigation it appears that in contrast to the ongoing experiment, considering the possibility to use lasers close to the allowed $5s^2S_{1/2} \rightarrow 5p\,^2P_{1/2,3/2}$ Rb transition, would probably be necessary to reach this objective.

\section*{Acknowledgements}
The authors are grateful to Alessio Ciamei, Florian Schreck, and the Amsterdam group, for providing us with experimental results prior to publication. A.D. and O.D. acknowledge the support of the ANR BLUESHIELD (Grant No. ANR-ANR-14-CE34-0006).

\bibliographystyle{unsrt}
\bibliography{RbSr_article}
\vfill\eject

\section*{Appendix}

\begin{figure}[H]
	\centering
	\includegraphics[width=8cm]{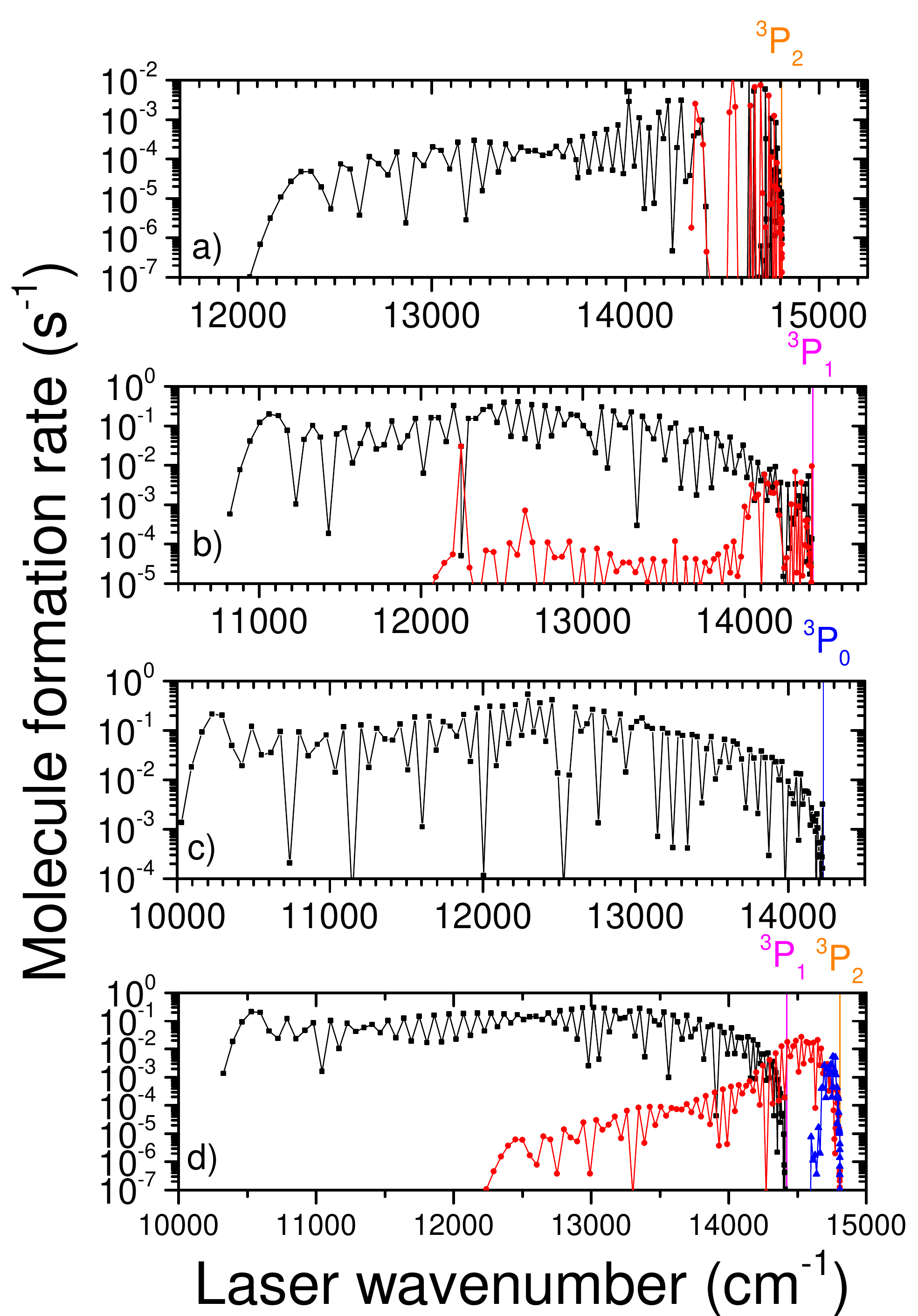}
	\caption{Ultracold molecule formation rates for $^{87}$Rb$^{84}$Sr levels as a function of the PA laser wavenumber, for states correlated to the $^{87}$Rb$(5s\,^2S)$+$^{84}$Sr($5s5p\,^3P_{0,1,2}$) dissociation limit. (a) (7) $\Omega=\frac{1}{2}$ in black, and (8) $\Omega=\frac{1}{2}$ in red. (b) (5) $\Omega=\frac{1}{2}$ in black, and (6) $\Omega=\frac{1}{2}$ in red. (c) (4) $\Omega=\frac{1}{2}$ in black. (d) (2) $\Omega=\frac{3}{2}$ in black, (3) $\Omega=\frac{3}{2}$ in red, (4) $\Omega=\frac{3}{2}$ in blue.}
	\label{fig:forbiddenUMF}
\end{figure}

\begin{figure}[H]
	\centering
	\includegraphics[width=12cm]{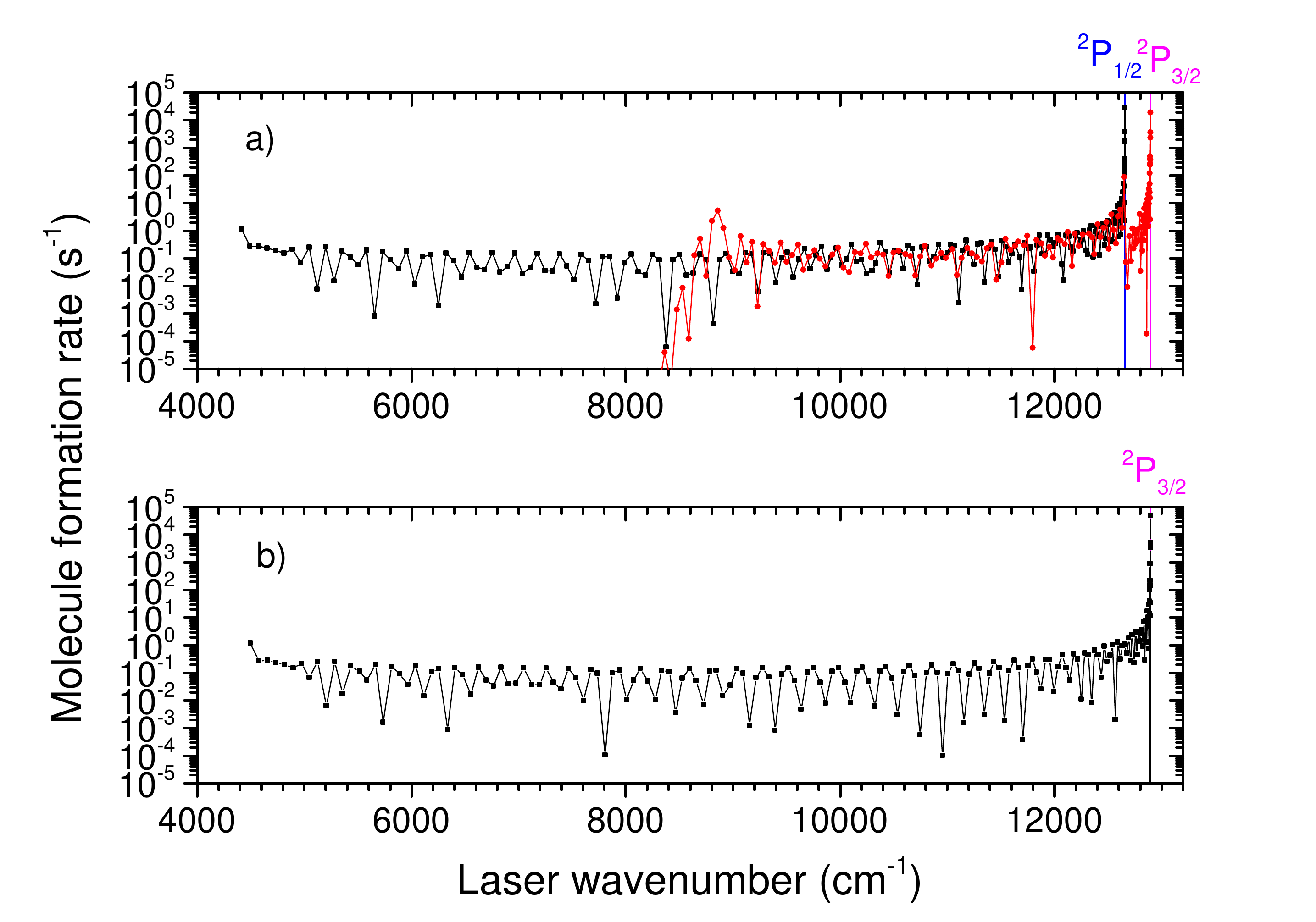}
	\caption{Ultracold molecule formation rates for the bound levels of (a) the (2) $\Omega=\frac{1}{2}$, (3) $\Omega=\frac{1}{2}$ states, and (b) of the (1) $\Omega=\frac{3}{2}$ state, correlated to the Rb$(5p\,^2P_{1/2,3/2})$+Sr($5s^2\,^1S$) dissociation limit.}
	\label{fig:allowedUMF}
\end{figure}

In Figs \ref{fig:forbiddenUMF} and \ref{fig:allowedUMF} we display the UMF rates corresponding to the situations treated in Section \ref{sec:PA}. We recall that these rates are obtained with Eq.(\ref{eq:CMrate}) which disregards the $R$-dependent TDM for the RE step. Therefore these graphs are intended to yield a global illustration of the variation of the UMF rate with the detunings, while the calculation of the vibrational distributions of the ground-state molecules are indeed computed taking in account these TDMs (see Eq. (\ref{eq:vibdist})). 

\end{document}